\documentclass[journal]{IEEEtran}
\usepackage{subfigure}
\usepackage{color}
\usepackage{cite}
\usepackage{amsmath}
\usepackage{cite}
\usepackage{graphicx}
\usepackage{epstopdf}
\usepackage{amsfonts,amsmath,amssymb}
\usepackage{graphicx}
\usepackage{url}
\usepackage{bm}
\usepackage{bbm}
\usepackage{subfigure}
\usepackage{stfloats}
\usepackage{color}

\usepackage{cite}

\DeclareMathOperator*{\argmax}{argmax}
\DeclareMathOperator*{\argmin}{argmin}

\newcommand{\Hnull}{\mathcal{H}_0}
\newcommand{\Halt}{\mathcal{H}_1}

\newcommand{\Honull}{\mathcal{{D}}_0}
\newcommand{\Hoalt}{\mathcal{{D}}_1}

\usepackage{cite,graphicx,amsmath,amssymb,cite,algorithm}

\newtheorem{corollary}{\textbf{Corollary}}
\newtheorem{theorem}{\textbf{Theorem}}

\newtheorem{remark}{\textbf{Remark}}

\begin{document}

\title{Covert Communication Achieved by A Greedy Relay in Wireless Networks}
\author{

{Jinsong Hu, Shihao Yan, Xiangyun Zhou, Feng Shu, Jun Li, and Jiangzhou Wang}

\thanks{This work was supported in part by the National Natural Science Foundation of China under Grant 61771244, Grant 61727802, Grant 61501238, and Grant 61472190, in part by the Australian Research Council's Discovery Projects (DP150103905), in part by the Open Research Fund of National Key Laboratory of Electromagnetic Environment, China Research Institute of Radiowave Propagation under Grant 201500013, in part by the Open Research Fund of the National Mobile Communications Research Laboratory, Southeast University, China, under Grant 2017D04, in part by the Jiangsu Provincial Science Foundation Project under Grant BK20150786, in part by the Specially Appointed Professor Program in Jiangsu Province, 2015, in part by the Fundamental Research Funds for the Central Universities under Grant 30916011205. This paper was presented in part at IEEE Global Communication Conference (GLOBECOM 2017), Singapore, Dec. 2017 \cite{Hu2017Covert}.
}

\thanks{J. Hu is with the School of Electronic and Optical Engineering, Nanjing University of Science and Technology, Nanjing, China, and also with the Research School of Engineering, Australian National University, Canberra, ACT, Australia (e-mail: jinsong\_hu@njust.edu.cn).}

\thanks{S. Yan is with the School of Engineering, Macquarie University, Sydney, NSW, Australia (e-mail: shihao.yan@mq.edu.au).}

\thanks{X. Zhou is with the Research School of Engineering, Australian National University, Canberra, ACT, Australia (e-mail: xiangyun.zhou@anu.edu.au).}

\thanks{F. Shu is with the School of Electronic and Optical Engineering, Nanjing University of Science and Technology, Nanjing, China, and also with the College of Computer and Information Sciences, Fujian Agriculture and Forestry University, Fuzhou, China (e-mail: shufeng@njust.edu.cn). }

\thanks{J. Li is with the School of Electronic and Optical Engineering, Nanjing University of Science and Technology, Nanjing, China,  also with the National Mobile Communications Research Laboratory, Southeast University, Nanjing, China, and also with the Department of Software Engineering, Institute of Cybernetics, National Research Tomsk Polytechnic University, Tomsk, Russia (e-mail: jun.li@njust.edu.cn). }

\thanks{J. Wang is with the School of Engineering and Digital Arts, University of Kent, Canterbury CT2 7NT, U.K. (e-mail: j.z.wang@kent.ac.uk).}

}
\maketitle

\vspace{-2cm}

\begin{abstract}
Covert wireless communication aims to hide the very existence of wireless transmissions in order to guarantee a strong security in wireless networks. In this work, we examine the possibility and achievable performance of covert communication in amplify-and-forward one-way relay networks. Specifically, the relay is greedy and opportunistically transmits its own information to the destination covertly on top of forwarding the source's message, while the source tries to detect this covert transmission to discover the illegitimate usage of the resource (e.g., power, spectrum) allocated only for the purpose of forwarding the source's information. We propose two strategies for the relay to transmit its covert information, namely rate-control and power-control the transmission schemes, for which the source's detection limits are analysed in terms of detection error probability and the achievable effective covert rates from the relay to destination are derived. Our examination determines the conditions under which the rate-control transmission scheme outperforms the power-control transmission scheme, and vice versa, which enables the relay to achieve the maximum effective covert rate. Our analysis indicates that the relay has to forward the source's message to shield its covert transmission and the effective covert rate increases with its forwarding ability (e.g., its maximum transmit power).
\end{abstract}

%
%
\section{Introduction}

\subsection{Background and Related Works}

Security and privacy are critical in existing and future wireless networks since a large amount of confidential information (e.g., credit card information, physiological information for e-health) is transferred over the open wireless medium \cite{bloch2011physical,zhou2011physical,yangnan2015}.
Traditional security techniques offer protection against eavesdropping through encryption, guaranteeing the integrity of messages over the air \cite{Menezes1996Handbook,Talbot2007Complexity}. However, it has been shown in the recent years that even the most robust encryption techniques can be defeated by a determined adversary. Physical-layer security, on the other hand, exploits the dynamic characteristics of the wireless medium to minimize the information obtained by eavesdroppers \cite{yan2016location,deng2016artificial,yan2016artificial,hu2017artificial,Liu2017Enhancing}. However, it does not provide protection against the detection of a transmission in the first place, which can offer an even stronger level of security, as the transmission of encrypted transmission can spark suspicion in the first place and invite further probing by skeptical eavesdroppers. On the contrary, apart from protecting the content of communication, covert communication aims to enable a wireless transmission between two users while guaranteeing a negligible detection probability of this transmission at a warden and thus achieving privacy of the transmitter.
Meanwhile, this strong security (i.e., hiding the wireless transmission) is desired in many application scenarios of wireless communications, such as covert military operations, location tracking in vehicular ad hoc networks and intercommunication of sensor networks or Internet of Things (IoT). Due to the broadcast nature of wireless channels, the security and privacy of wireless communications has been an ever-increasing concern, which now is the biggest barrier to the wide-spread adoption of sensor networks or IoT technologies \cite{Mukherjee2015Physical}. In sensor networks or IoT, multiple hidden transmitters or receivers, which may be surrounded or monitored by wardens/cybercriminals, are trying to exchange critical information through multi-hop wireless transmissions. Each transmission should be kept covert to enable the end-to-end covert communication in order to guarantee the `invisibility' of the transmitters. As such, the hiding of communication termed covert communication or low probability of detection communication, which aims to shield the very existence of wireless transmissions against a warden to achieve security, has recently drawn significant research interests and is emerging as a cutting-edge technique in the context of wireless communication security\cite{Pak2014reliable,bash2015hiding,bloch2016covert}.

Although spread-spectrum techniques are widely used to achieve covertness in military applications of wireless communications \cite{Simon1994Spread}, many fundamental problems have not been well addressed. This leads to the fact that the probability that the spread-spectrum techniques fail to hide wireless transmissions is unknown, significantly limiting its application.
The fundamental limit of covert communication has been studied under various channel conditions, such as additive white Gaussian noise (AWGN) channels \cite{bash2013limits}, binary symmetric channels \cite{pak2013reliable}, discrete memoryless channels \cite{wang2016fundamental}, and multiple input multiple output (MIMO) AWGN channels \cite{Abdelaziz2017Fundamental}. It is proved that $\mathcal{O}(\sqrt{n})$ bits of information can be transmitted to a legitimate receiver reliably and covertly in $n$ channel uses as $n \rightarrow \infty$. This means that the associated covert rate is zero due to $\lim_{n\rightarrow\infty}\mathcal{O}(\sqrt{n})/n\rightarrow0$. Following these pioneering works on covert communication, a positive rate has been proved to be achievable when the warden has uncertainty on his receiver noise power \cite{lee2015achieving,BiaoHe2017on}, or an uniformed jammer comes in to help \cite{Sobers2017Covert}. Most recently, \cite{goeckel2016covert} has examined the impact of noise uncertainty on covert communication. In addition, the effect of the imperfect channel state information (CSI) and finite blocklength (i.e., finite $n$) on covert communication has been investigated in \cite{Shahzad2017Covert} and \cite{ShihaoYan2017Covert}, respectively.

\subsection{Motivation and Our Contributions}

The ultimate goal of covert wireless communication is to establish shadow wireless networks \cite{bash2015hiding}, in which each hop transmission should be kept covert to enable the end-to-end covert communication, in order to guarantee the ``invisibility" of the transmitters. Following the previous works that only focused on covert transmissions in point-to-point communication scenarios, in this work, for the first time, we consider covert communications in the context of amplify-and-forward one-way relay networks. This is motivated by the scenario where the relay (R) tries to transmit its own information to the destination (D) on top of forwarding the information from the source (S) to D.
Specifically, for example, in some relay networks (possible application scenarios of sensor networks or IoT) the communication resources (e.g., spectrum, power) can be managed or owned by S, where S may not allow R to transmit its own information on top of forwarding S's messages to D. This is due to the fact that R's additional transmission may cause interference within the specific spectrum owned/managed by S and also consume more transmit power, which is possibly wirelessly transferred from S (owned by S) and should be only used for forwarding S's information. Therefore, this additional transmission of R should be kept covert from S.

We note that conceptually the covert transmission from R to D is similar to steganography, in which covert information is transmitted by hiding in innocuous objects \cite{Provos2003hide}. These innocuous objects are utilized as ``cover text" to carry the covert information. In our work, the innocuous objects are the forwarding transmissions from R to D. The main difference between our work and steganography is that in our work the covert information is shielded by the forwarding transmissions from R to D at the physical layer, while in steganography the covert information is hidden and transmitted by encoding or modifying some contents (e.g., shared videos or images) at the application layer (as discussed in Section~III of \cite{bash2015hiding}).

In the literature, covert communications with positive transmission rate are achieved in the context of point-to-point systems by considering different uncertainty sources, such as random received noise power \cite{BiaoHe2017on}, random jamming signals \cite{Sobers2017Covert}, and imperfect CSI \cite{Shahzad2017Covert}. In the considered relay networks, as mentioned above the uncertainty is inherently embedded in the forwarding strategies of the S's information from R to D, where the covert transmission with a positive rate from R to D does not require any extra uncertainty sources.
The performance of the considered covert communication in relay networks and the covert communication in other point-to-point communication systems highly depends on the amount of uncertainty appeared in the system model. As such, it is hard to compare the achieved covert rate limits or warden's detection limits directly, since the uncertainty sources are different and it is hard to quantify the corresponding amount of uncertainty in the same manner.

Our main contributions are summarized below.
\begin{itemize}
\item We examine the possibility and achievable performance of covert communications in one-way relay networks. Specifically, we propose two strategies for R to transmit the covert information to D, namely the rate-control and power-control transmission schemes, in which the transmission rate and transmit power of the covert message are fixed and to be optimized regardless of the channel quality from R to D, respectively. We also identify the necessary conditions that the covert transmission from R to D can possibly occur without being detected by S with probability one and clarify how R hides its covert transmission in forwarding S's message to D in these two schemes.

\item We derive the detection limits at S in terms of the prior probability of null hypothesis $1-\omega$, the prior probability of alternative hypothesis $\omega$, the false alarm rate $\alpha$ and miss detection rate $\beta$ are in closed-from expressions for the proposed two transmission schemes. Then, we determine the optimal detection threshold at S, which minimizes the detection error probability $\xi = (1-\omega)\alpha + \omega\beta$ and obtain the associated minimum detection error probability $\xi^{\ast}$. Our analysis leads to many useful insights. For example, we analytically prove that $\xi^{\ast}$ increases with R's maximum transmit power, which indicates that boosting the forwarding ability of R also increases its capacity to perform covert transmissions. This demonstrates a tradeoff between the achievable effective covert rate and R's ability to aid the transmission from S to D.

\item We analyze the effective covert rates achieved by these two schemes subject to the covert constraint $\xi^{\ast} \geq \min(1-\omega,\omega) - \epsilon$, where $\epsilon \in [0,1]$ is predetermined to specify the covert constraint. Our analysis indicates that the achievable effective covert rate approaches zero as the transmission rate from S to D approaches zero, which demonstrates that covert transmission from R to D is only feasible with the legitimate transmission from S to D as the shield. Our examination shows that the rate-control transmission scheme outperforms the power-control transmission scheme under some specific conditions, and vise versa. Our examination enables R to switch between these two schemes in order to achieve a higher effective covert rate.
\end{itemize}

The rest of this paper is organized as follows. Section II details our system model and adopted assumptions. Section III and IV present the rate-control and power-control transmission schemes, respectively. Thorough analysis on the performance of these two transmission are provided in these two sections as well. Section V provides numerical results to confirm our analysis and provide useful insights on the impact of some parameters. Section VI draws conclusions.

\emph{Notation:} Scalar variables are denoted by italic symbols. Vectors is denoted by lower-case boldface symbols. Given a complex number, $|\cdot|$ denotes the modulus. Given a complex vector, $(\cdot)^{\dag}$ denotes the conjugate transpose. $\mathbb{E}[\cdot]$ denotes expectation operation.

\section{System Model}

\subsection{Considered Scenario and Adopted Assumptions}

\begin{figure}[!t]
    \begin{center}
        \includegraphics[width=0.98\columnwidth]{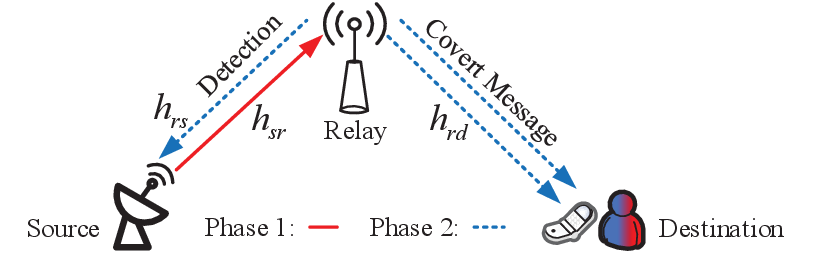}
        \caption{Covert communication in one-way relay networks.}\label{fig1}
    \end{center}
\end{figure}

As shown in Fig.~\ref{fig1}, in this work we consider a one-way relay network, in which S transmits information to D with the aid of R, since a direct link from S to D is not available. As mentioned in the introduction, S allocates some resource to R in order to seek its help to relay the message to D. However, in some scenarios R may intend to use this resource to transmit its own message to D as well, which is forbidden by S and thus should be kept covert from S. As such, in the considered system model S is also the warden to detect whether R transmits its own information to D when it is aiding the transmission from S to D.

We assume the wireless channels within our system model are subject to independent quasi-static Rayleigh fading with equal block length and the channel coefficients are independent and identically distributed (i.i.d.) circularly symmetric complex Gaussian random variables with zero-mean and unit-variance. We also assume that each node is equipped with a single antenna. The channel from S to R is denoted by $h_{sr}$ and the channel from R to D is denoted by $h_{rd}$. We assume R knows both $h_{sr}$ and $h_{rd}$ perfectly, while S only knows $h_{sr}$ and D only knows $h_{rd}$. Considering channel reciprocity, we assume the channel from R to S (denoted by $h_{rs}$) is the same as $h_{sr}$ and thus it is perfectly known by S. We further assume that R operates in the half-duplex mode and thus the transmission from S to D occurs in two phases: phase 1 (S transmits to R) and phase 2 (R transmits to D).

\subsection{Transmission from Source to Relay (Phase 1)}

In phase 1, the received signal at R is given by
\begin{align}
\mathbf{y}_r[i]=\sqrt{P_s}h_{sr}\mathbf{x}_b[i]+\mathbf{n}_r[i],
\end{align}
where $P_s$ is the fixed transmit power of S, $\mathbf{x}_b$ is the transmitted signal by S satisfying $\mathbb{E}[\mathbf{x}_b[i]\mathbf{x}^{\dag}_b[i]]=1$, $i = 1, 2, \dots, n$ is the index of each channel use ($n$ is the total number of channel uses in each phase), and $\mathbf{n}_r[i]$ is the AWGN at relay with $\sigma^2_r$ as its variance, i.e., $\mathbf{n}_r[i] \thicksim\mathcal{CN}(0,\sigma^2_r)$.
In the literature, multiple approaches have been developed to estimate the noise variance at a receiver. In general, these approaches can be divided into two major categories: data-aided (DA) approaches and non-data-aided (NDA) approaches~\cite{Cui2006Power}. The DA approaches assume that transmitted symbols are known at the receiver and maximum-likelihood estimation can be used to estimate the noise variance. For the NDA approaches, transmit symbols are unknown at the receiver and the noise variance is based on the statistics of the received signals.
In this work, we consider that R operates in the AF mode and thus R will forward a linearly amplified version of the received signal to D in phase 2. As such, the forwarded (transmitted) signal by R is given by
\begin{align}\label{x_r}
\mathbf{x}_r[i]=G\mathbf{y}_r[i]=G(\sqrt{P_s}h_{sr}\mathbf{x}_b[i]+\mathbf{n}_r[i]),
\end{align}
which is a linear scaled version of the received signal by a scalar $G$. In order to guarantee the power constraint at R, the value of $G$ is chosen such that $\mathbb{E}[\mathbf{x}_r[i]\mathbf{x}^{\dag}_r[i]]=1$, which leads to $G=1/\sqrt{P_s|h_{sr}|^2+\sigma^2_r}$.

In this work, we also consider that the transmission rate from S to D is predetermined, which is denoted by $R_{sd}$. We also consider a maximum power constraint at R, i.e., $P_r \leq P_r^{\mathrm{max}}$. As such, although R knows both $h_{sr}$ and $h_{rd}$ perfectly, transmission outage from S to D still incurs when $C_{sd}^{\mathrm{max}}<R_{sd}$, where $C_{sd}^{\mathrm{max}}$ is the channel capacity from S to D for $P_r = P_r^{\mathrm{max}}$. Then, the transmission outage probability is given by $\delta = \mathcal{P}[C_{sd}^{\mathrm{max}}<R_{sd}]$, which has been derived in a closed-form expression \cite{Molu2013an}. We assume that all the nodes in the network do not transmit signals when the outage occurs. In practice, the pair of $R_{sd}$ and $\delta$ determines the specific aid (i.e., the value of $P_r^{\mathrm{max}}$) required by S from R, which relates to the amount of resource allocated to R by S as a payback. In this work, we assume both $R_{sd}$ and $\delta$ are predetermined, which leads to a predetermined $P_r^{\mathrm{max}}$.

\subsection{Transmission Strategies at Relay (Phase 2)}

In this subsection, we detail the transmission strategies of R when it does and does not transmit its own information to D. We also determine the condition that R can transmit its own message to D without being detected by S with probability one, in which the probability to guarantee this condition is also derived.

\subsubsection{Relay's Transmission without the Covert Message}

In the case when the relay does not transmit its own message (i.e., covert message) to D, it only transmits $\mathbf{x}_r$ to D. Accordingly, the received signal at D is given by
\begin{align}
&\mathbf{y}_d[i]=\sqrt{P_r^0}h_{rd}\mathbf{x}_r[i]+\mathbf{n}_d[i]  \notag \\ &=\sqrt{P_r^0}Gh_{rd}\sqrt{P_s}h_{sr}\mathbf{x}_b[i]+\sqrt{P_r^0}Gh_{rd}\mathbf{n}_r[i]+\mathbf{n}_d[i],
\end{align}
where $P_r^0$ is the transmit power of $\mathbf{x}_r$ at R in this case and $\mathbf{n}_d[i]$ is the AWGN at D with $\sigma^2_d$ as its variance, i.e., $\mathbf{n}_d[i]\thicksim\mathcal{CN}(0,\sigma^2_d)$. Accordingly, the signal-to-noise ratio (SNR) at the destination for $\mathbf{x}_b$, which has been derived in a closed-form expression in~\cite{Deng2015full}, is given by
\begin{align} \label{gamma_d1}
\gamma_d&=\frac{P_s|h_{sr}|^2P_r^0|h_{rd}|^2 G^2}{P_r^0|h_{rd}|^2G^2\sigma^2_r+\sigma^2_d} \notag \\ &=\frac{\gamma_{1}\gamma_{2}}{\gamma_{1}+\gamma_{2}+1},
\end{align}
where $\gamma_{1}\triangleq(P_s|h_{sr}|^2)/\sigma^2_r$, $\gamma_{2}\triangleq(P_r^0|h_{rd}|^2)/\sigma^2_d$, and the scalar $G$ is defined earlier as $G=1/\sqrt{P_s|h_{sr}|^2+\sigma^2_r}$.

For a predetermined $R_{sd}$, R does not have to adopt the maximum transmit power for each channel realization in order to guarantee a specific transmission outage probability. When the transmission outage occurs (i.e., $C_{sd}^{\mathrm{max}}<R_{sd}$ occurs), R will not transmit (i.e., $P_r^0 = 0$). When $C_{sd}^{\mathrm{max}} \geq R_{sd}$, R only has to ensure $C_{sd} = R_{sd}$, where $C_{sd} = {1}/{2}\log_2(1+\gamma_d)$. Then, following \eqref{gamma_d1} the transmit power of R when $C_{sd}^{\mathrm{max}} \geq R_{sd}$ is given by $P_r^{0}={\mu\sigma^2_d}/{|h_{rd}|^2}$, where
\begin{align}\label{mu}
\mu\triangleq\frac{(P_s|h_{sr}|^2+\sigma^2_r)(2^{2R_{sd}}-1)}{\left[P_s|h_{sr}|^2-\sigma^2_r(2^{2R_{sd}}-1)\right]}.
\end{align}
We note that \eqref{mu} indicates that R does not use its maximum transmit power $P_r^{\mathrm{max}}$ to forward S's information when it does not transmit covert information to D. This is due to the fact that the transmission from S to D is of a fixed rate $R_{sd}$ and a larger transmit power that leads to $C_{sd} > R_{sd}$ (not $C_{sd} = R_{sd}$) does not bring in extra benefit to this transmission from S to D. As such, in order to save energy R only sets its transmit power as per \eqref{mu} to guarantee $C_{sd} = R_{sd}$.
Noting $\gamma_d < \gamma_1$, we have $1/2\log_2(1+\gamma_1)>R_{sd}$ when $C_{sd} = R_{sd}$. As such, $\mu$ given in \eqref{mu} is nonnegative.
Following \eqref{gamma_d1}, we note that $C_{sd}^{*} \geq R_{sd}$ requires $|h_{rd}|^2 \geq {\mu\sigma^2_d}/{P_r^{\mathrm{max}}}$. As such, the transmit power of R without a covert message is given by
\begin{align}\label{PR0}
P_r^{0}=\left\{
  \begin{array}{ll}
    \frac{\mu\sigma^2_d}{|h_{rd}|^2}, & |h_{rd}|^2 \geq \frac{\mu\sigma^2_d}{P_r^{\mathrm{max}}}, \\
    0,  &|h_{rd}|^2 <\frac{\mu\sigma^2_d}{P_r^{\mathrm{max}}}.
  \end{array}
\right.
\end{align}
As per \eqref{PR0}, we note that relay will forward message when $|h_{rd}|^2 \geq\mu\sigma^2_d/P_r^{\mathrm{max}}$ is met. We denote this necessary condition as $\mathbb{B}$. As such, R will forward $\mathbf{x}_b$ to D and S will perform detection whenever condition $\mathbb{B}$ is met. Considering quasi-static Rayleigh fading, the cumulative distribution function (cdf) of $|h_{rd}|^2$ is given by $F_{|h_{rd}|^2}(x) = 1 - e^{-x}$ and thus the probability that $\mathbb{B}$ is guaranteed is given by
\begin{align}\label{P_b}
\mathcal{P}_B = \exp\left\{-\frac{\mu\sigma^2_d}{P_r^{\mathrm{max}}}\right\}.
\end{align}

\subsubsection{Relay's Transmission with the Covert Message}

In the case when R transmits the covert message to D on top of forwarding $\mathbf{x}_b$, the received signal at D is given by
\begin{align}\label{YD2}
\mathbf{y}_d[i]&=\sqrt{P_r^1}Gh_{rd}\sqrt{P_s}h_{sr}\mathbf{x}_b[i]+\sqrt{P_{\Delta}}h_{rd}\mathbf{x}_c[i]+ \notag \\
&~~~\sqrt{P_r^1}Gh_{rd}\mathbf{n}_r[i]+\mathbf{n}_d[i],
\end{align}
where $P_r^1$ is R's transmit power of $\mathbf{x}_b$ in this case and $P_{\Delta}$ is R's transmit power of the covert message $\mathbf{x}_c$ satisfying $\mathbb{E}[\mathbf{x}_c[i]\mathbf{x}^{\dag}_c[i]]=1$.
We note that the covert transmission from R to D should not affect the transmission from S to D. Otherwise, S can easily observe this covert transmission. As such, here we assume D always first decodes $\mathbf{x}_b$ with $\mathbf{x}_c$ as interference. Following \eqref{YD2}, the signal-to-interference-plus-noise ratio (SINR) for $\mathbf{x}_b$ is given by
\begin{align}\label{gamma_d2}
\gamma_d&=\frac{P_s|h_{sr}|^2P_r^1|h_{rd}|^2 G^2}{P_r^1|h_{rd}|^2G^2\sigma^2_r+P_{\Delta}|h_{rd}|^2+\sigma^2_d} \notag \\
&=\frac{\gamma_{1}\gamma_{3}}{\gamma_{3}+(\gamma_{1}+1)\left(\gamma_{3}P_{\Delta}/P_r^1+1\right)},
\end{align}
where $\gamma_{3}\triangleq(P_r^1|h_{rd}|^2)/\sigma^2_d$. We will determine $P_r^{1}$ based on different transmission strategies of the covert message from R to D.

\subsection{Decoding of the Covert Message}
As discussed above, the covert transmission from R to D should not affect the transmission from S to D and thus we have to guarantee the successful decoding of $\mathbf{x}_b$ even when $\mathbf{x}_c$ is treated as interference to $\mathbf{x}_b$. We also note that this covert transmission cannot happen when the transmission outage from S to D occurs.  This is, for example, due to the fact that when the transmission outage occurs R will request a retransmission from S, which enables S to detect R's covert transmission with probability one if the covert transmission happened. Therefore, the covert transmission from R to D only occur when the successful transmission from S to D is guaranteed (i.e., when $\mathbf{x}_b$ is successfully decoded at D).
As such, when the covert message is transmitted by R, successive interference cancellation (SIC) that allows a receiver to decode different signals that arrive simultaneously is implemented at D. Taking advantage of SIC, D decodes the stronger signal (i.e., $\mathbf{x}_b$) first, subtracts it from the combined signal $\mathbf{y}_d$ given in \eqref{YD2}, and finally decodes the weaker one (i.e., $\mathbf{x}_c$) from the residue.
We also note that D cannot jointly decode $\mathbf{x}_b$ and $\mathbf{x}_c$ due to the fact that the codebooks used for encoding $\mathbf{x}_b$ and $\mathbf{x}_c$ are different in order to guarantee that the codebook for $\mathbf{x}_c$ is unknown while the codebook for $\mathbf{x}_b$ is known to the S.
Hence, the effective received signal used to decode the covert message $\mathbf{x}_c$ is given by
\begin{align} \label{tilde y_d}
\tilde{\mathbf{y}}_d[i]=\sqrt{P_{\Delta}}h_{rd} \mathbf{x}_c[i]+\sqrt{P_r^1}h_{rd} G \mathbf{n}_r[i] + \mathbf{n}_d[i].
\end{align}
Then, following \eqref{tilde y_d} the SINR for $\mathbf{x}_c$ is
\begin{align} \label{gamma_c}
\gamma_{\Delta} = \frac{P_{\Delta}|h_{rd}|^2}{P_r^1|h_{rd}|^2 G^2 \sigma^2_r +\sigma^2_d}.
\end{align}

\subsection{Binary Detection at Source and the Covert Constraint}

In this subsection, we present the optimal detection strategy adopted by S (i.e., Source).

In phase 2 when R transmits to D, S will detect whether R transmits the covert message $\mathbf{x}_c$ on top of forwarding S's message $\mathbf{x}_b$ to D. R does not transmit $\mathbf{x}_c$ in the null hypothesis $\Hnull$ while it does in the alternative hypothesis $\Halt$. Then, the received signal at S in phase 2 is given by
\begin{align} \label{yw}
\mathbf{y}_s[i]&=\left\{
  \begin{array}{ll}
    \sqrt{P_r^0}h_{rs}\mathbf{x}_r[i] \!+\! \mathbf{n}_s[i], &~~~~~~\Hnull, \\
    \sqrt{P_r^1}h_{rs}\mathbf{x}_r[i] \!+\! \sqrt{P_{\Delta}}h_{rs}\mathbf{x}_c[i] \!+\! \mathbf{n}_s[i],  &~~~~~~\Halt,
  \end{array}
\right. \notag \\
&=\left\{
  \begin{array}{ll}
    \frac{\sqrt{P_r^0}h_{rs}}{\sqrt{P_s|h_{sr}|^2\!+\!\sigma^2_r}}(\sqrt{P_s}h_{sr}\mathbf{x}_b[i]\!+\!\mathbf{n}_r[i]) \!+\! \mathbf{n}_s[i], &\Hnull, \\
    \frac{\sqrt{P_r^1}h_{rs}}{\sqrt{P_s|h_{sr}|^2\!+\!\sigma^2_r}}(\sqrt{P_s}h_{sr}\mathbf{x}_b[i]+ \\
    \mathbf{n}_r[i])\!+\! \sqrt{P_{\Delta}}h_{rs}\mathbf{x}_c[i] \!+\! \mathbf{n}_s[i],  &\Halt.
  \end{array}
\right.
\end{align}
Noting that $\mathbf{x}_b[i]$ is known by S, hence, S can cancel the corresponding component from its received signal $\mathbf{y}_s[i]$, due to the fact the infinite blocklength is considered in this work and S can exactly estimate the scale factor of $\mathbf{x}_b[i]$. Then, the effective received signal used for detection at S is given by
\begin{align}\label{yw_tilde}
\tilde{\mathbf{y}}_s[i]\!=\!
\left\{
  \begin{array}{ll}
    \frac{\sqrt{P_r^0}h_{rs}}{\sqrt{P_s|h_{sr}|^2\!+\!\sigma^2_r}}\mathbf{n}_r[i] \!+\! \mathbf{n}_s[i], &\Hnull, \\
    \frac{\sqrt{P_r^1}h_{rs}}{\sqrt{P_s|h_{sr}|^2\!+\!\sigma^2_r}}\mathbf{n}_r[i]\!+\! \sqrt{P_{\Delta}}h_{rs}\mathbf{x}_c[i] \!+\! \mathbf{n}_s[i],  &\Halt.
  \end{array}
\right.
\end{align}
Following \eqref{yw_tilde}, the probability density functions of the observations $\tilde{\mathbf{y}}_s$ under $\Hnull$ and $\Halt$ are, respectively, given by
\begin{align}\label{likelihood func H0}
f(\tilde{\mathbf{y}}_s\big{|}\Hnull)&=\prod\limits_{i=1}^n f(\tilde{\mathbf{y}}_s[i]\big{|}\Hnull) \notag \\
&=\frac{1}{\left(2\pi \sigma_{\Hnull}^2\right)^{\frac{n}{2}}}\exp\left\{-\frac{1}{2 \sigma_{\Hnull}^2}\sum\limits_{i=1}^n |\tilde{\mathbf{y}}_s[i]|^2\right\},
\end{align}
\begin{align}\label{likelihood func H1}
f(\tilde{\mathbf{y}}_s\big{|}\Halt)&=\prod\limits_{i=1}^n f(\tilde{\mathbf{y}}_s[i]\big{|}\Halt) \notag \\
&=\frac{1}{\left(2\pi \sigma_{\Halt}^2\right)^{\frac{n}{2}}}\exp\left\{-\frac{1}{2 \sigma_{\Halt}^2}\sum\limits_{i=1}^n |\tilde{\mathbf{y}}_s[i]|^2\right\},
\end{align}
where $\sigma_{\Hnull}^2\triangleq P_r^0 |h_{rs}|^2\sigma_r^2/(P_s|h_{sr}|^2\!+\!\sigma_r^2)\!+\! \sigma_s^2$ and $\sigma_{\Halt}^2\triangleq P_r^1 |h_{rs}|^2\sigma_r^2/(P_s|h_{sr}|^2\!+\!\sigma_r^2) \!+\! P_{\Delta}|h_{rs}|^2\!+\!\sigma_s^2$. Following \eqref{likelihood func H0} and \eqref{likelihood func H1}, based on the Fisher-Neyman factorization theorem~\cite{DeGroot2011Probability}, we note that the term $T(n)=\sum_{i=1}^{n}|\tilde{\mathbf{y}}_s[i]|^2$ is the sufficient test statistic for the detector at S. As such, the detector at S for an arbitrary threshold is given by
\begin{align}\label{decisions}
\frac{1}{n}T(n)\mathop{\gtrless}\limits_{\Honull}^{\Hoalt}\tau,
\end{align}
where $\tau$ is the threshold for (1/n)$T(n)$, which will be determined later, $\Hoalt$ and $\Honull$ are the binary decisions that infer whether R transmits covert message or not, respectively. We will examine how S sets the optimal value of $\tau$ in order to minimize the detection error probability in the following sections for considered different transmission strategies. Considering infinite blocklength, i.e., $n \rightarrow \infty$, we have
\begin{eqnarray}\label{Pw}
\lim_{n \rightarrow \infty}\frac{1}{n}T(n)\!=\!
 \left\{ \begin{aligned}\label{T}
        \ &P_r^0 |h_{rs}|^2\phi + \sigma_s^2, ~~~~~~~~~~~~~~~~\Hnull, \\
        \ &P_r^1 |h_{rs}|^2\phi + P_{\Delta}|h_{rs}|^2 \!+\! \sigma_s^2,  ~~~\Halt,
         \end{aligned} \right.
\end{eqnarray}
where $\phi\triangleq \sigma_r^2/(P_s |h_{sr}|^2+\sigma_r^2)$.

The detection performance of S is normally measured by its detection error probability, which is defined as
\begin{align} \label{xi}
\xi\triangleq (1-\omega)\alpha  + \omega\beta,
\end{align}
where $\omega=\mathcal{P}(\Halt)$ is the probability that R transmits a covert message, $1-\omega=\mathcal{P}(\Hnull)$ is the probability that R does not transmit a covert message, $\alpha=\mathcal{P}(\Hoalt|\Hnull)$ is S's false alarm rate, and $\beta=\mathcal{P}(\Honull|\Halt)$ is S's miss detection rate.

In practice, it is hard to know $\xi$ at R since the threshold $\tau$ adopted by S is unknown. In this work, we consider the worst-case scenario where $\tau$ is optimized at S to minimize $\xi$. As such, the covert constraint considered in this work is $\xi^{\ast} \geq \min\{1-\omega, \omega\} - \epsilon$, where $\xi^{\ast}$ is the minimum detection error probability achieved at S.

\section{Rate-Control Transmission Scheme}
In this section, we consider the rate-control transmission scheme, in which R transmits a covert message to D with a constant rate when some specific realizations of $|h_{rd}|^2$ are guaranteed. To this end, R varies its transmit power as per $h_{rd}$ such that $P_{\Delta}|h_{rd}|^2$ is fixed as $Q$. Specifically, we first determine R's transmit power in $\Halt$ and then analyze the detection error probability at S, based on which we also derive S's optimal detection threshold. Furthermore, we derive the effective covert rate achieved by the rate-control transmission scheme.

\subsection{Transmit Power at Relay under $\Halt$}
Following \eqref{gamma_d2} and defining $Q = P_{\Delta}|h_{rd}|^2$, in order to guarantee $C_{sd} = R_{sd}$ under $\Halt$, $P_r^{1}$ is given as
\begin{align}\label{Pr1_Q1}
P_r^{1}=\frac{\mu(Q+\sigma^2_d)}{|h_{rd}|^2},
\end{align}
which requires $C_{sd}^{*} \geq R_{sd}$ that leads to $ |h_{rd}|^2 \geq (\mu\sigma^2_d+\mu Q+Q)/P_r^{\mathrm{max}}$.
We note that $P_r^{1}$ is the transmit power of the relay to forward the signal from S to D. In practical scenario, R can set the value of $P_r^{1}$ as per the system parameters $h_{sr}$, $h_{rd}$, $P_s$, $\sigma_r^2$, $\sigma_d^2$, $R_{sd}$, and $Q$. The values of these system parameters are known by R. Specifically, $h_{sr}$ can $h_{rd}$ can be obtained through channel estimations. The values of $\sigma_r^2$ and $\sigma_d^2$ can be achieved through \emph{a priori} measurements collected from the environment, where $\sigma_d^2$ is fed back from D to R. The value of $R_{sd}$ is predetermined by the QoS requirement of the communication from S to D, while the value of $Q$ is a design parameter to determine at R.
Considering the maximum power constraint at R (i.e., $P_r^1 + P_{\Delta} \leq P_r^{\mathrm{max}}$ under this case), R has to give up the transmission of the covert message (i.e., $P_{\Delta} = 0$) when $P_r^1>P_r^{\mathrm{max}}-P_{\Delta}$ and sets $P_r^1$ the same as $P_r^0$ given in  \eqref{PR0}. This is due to the fact that S knows $h_{rs}$ and it can detect with probability one when the total transmit power of R is greater than $P_r^{\mathrm{max}}$. Then, the transmit power of $\mathbf{x}_r$ under $\Halt$ for the rate-control transmission scheme is given by
\begin{align}
P_r^{1}=\left\{
  \begin{array}{ll}
    \frac{\mu(Q+\sigma^2_d)}{|h_{rd}|^2}, & |h_{rd}|^2 \geq \frac{\mu\sigma^2_d+\mu Q+Q}{P_r^{\mathrm{max}}}, \\
    \frac{\mu\sigma^2_d}{|h_{rd}|^2}, & \frac{\mu\sigma^2_d}{P_r^{\mathrm{max}}} \leq |h_{rd}|^2 < \frac{\mu\sigma^2_d+\mu Q+Q}{P_r^{\mathrm{max}}}, \\
    0,  &|h_{rd}|^2 <\frac{\mu\sigma^2_d}{P_r^{\mathrm{max}}}.
  \end{array}
\right.  \label{Pr1_Q}
\end{align}
As per \eqref{Pr1_Q}, when R cannot support the transmission from S to D (i.e., when $|h_{rd}|^2 <{\mu\sigma^2_d}/{P_r^{\mathrm{max}}}$), R or D will send the retransmission request to S and R should not forward $\mathbf{x}_b$, since this forwarding will definitely fail. In the meantime, S is aware of that the received energy comes from the R's covert transmission if R has transmitted the covert message during this period. Due to that the CSI of all the channels is available to R, R knows exactly when the transmission outage from R to D occurs and thus R will not transmit covert information to D when this outage occurs.
In summary, S cannot detect R's covert transmission with probability one only when the condition $ |h_{rd}|^2 \geq (\mu\sigma^2_d+\mu Q+Q)/P_r^{\mathrm{max}}$ is guaranteed. We denote this necessary condition for covert communication as $\mathbb{C}$. Considering quasi-static Rayleigh fading, the cumulative distribution function (cdf) of $|h_{rd}|^2$ is given by $F_{|h_{rd}|^2}(x) = 1 - e^{-x}$ and thus the probability that $\mathbb{C}$ is guaranteed is given by
\begin{align}\label{P_c_Q}
\mathcal{P}_C = \exp\left\{-\frac{\mu\sigma^2_d+\mu Q+Q}{P_r^{\mathrm{max}}}\right\}.
\end{align}
We note that $\mathcal{P}_C$ is a monotonically decreasing function of $Q$, which indicates that the probability that R will transmit a covert message decreases as $Q$ increases.

In this work, we consider quasi-static Rayleigh fading channels where the channels remain constant within each transmission period and vary independently from one period to another. We would like to clarify that R could possibly transmit a covert message to D without being detected during a retransmission from S to D (i.e., new transmission period) when the condition $\mathbb{C}$ is met.

\subsection{Detection Error Probability at Source}

In this subsection, we derive S's false alarm rate, i.e., $\alpha$, and miss detection rate, i.e., $\beta$.

\begin{theorem}\label{theorem1}
When the condition $\mathbb{B}$ is guaranteed, for a given $\tau$, the false alarm and miss detection rates at S are derived as
\begin{align}
\alpha&=\left\{
  \begin{array}{ll}
    1,  &\tau<\sigma_s^2, \\
    1-\mathcal{P}_B^{-1}\kappa_1(\tau), & \sigma_s^2\leq\tau\leq\rho_1,\\
    0,  &\tau>\rho_1,
  \end{array}
\right. \label{PFA_Q}\\
\beta&=\left\{
  \begin{array}{ll}
    0,  &\tau<\sigma_s^2, \\
    \mathcal{P}_B^{-1}\kappa_2(\tau), & \sigma_s^2\leq\tau\leq \rho_2,\\
    1,  &\tau>\rho_2,
  \end{array}
\right. \label{PMD_Q}
\end{align}
where
\begin{align}
&\rho_1\triangleq P_r^{\mathrm{max}}|h_{rs}|^2\phi+\sigma_s^2, \label{rho1_ex} \\
&\rho_2\triangleq P_r^{\mathrm{max}}|h_{rs}|^2\left(\phi+\frac{(\phi\mu+1)Q}{\mu\sigma^2_d}\right)+\sigma_s^2, \notag \\
&\kappa_1(\tau)\triangleq\mathrm{exp}\left\{-\frac{\phi\mu\sigma^2_d|h_{rs}|^2}
    {\tau-\sigma_s^2}\right\}, \notag  \\
&\kappa_2(\tau)\triangleq\mathrm{exp}\left\{-\frac{\left(\phi\mu\sigma^2_d+\left(\phi\mu+1\right)Q\right)|h_{rs}|^2}{\tau-\sigma_s^2}\right\}. \notag
\end{align}
\end{theorem}
\begin{IEEEproof}
Considering the maximum power constraint at R under $\Hnull$ (i.e., $P_r^0 \leq P_r^{\mathrm{max}}$) and following \eqref{PR0}, \eqref{decisions}, and \eqref{Pw}, the false alarm rate under the condition $\mathbb{B}$ is given by
\begin{align}
&\alpha =\mathcal{P}\left[\frac{\mu\sigma^2_d}{|h_{rd}|^2}|h_{rs}|^2\phi+\sigma_s^2\geq\tau\big{|}\mathbb{B}\right]  \notag \\
&=\left\{
  \begin{array}{ll}
    1,  &\tau<\sigma_s^2, \\
    {\mathcal{P}\left[\frac{\mu\sigma^2_d}{P_r^{\mathrm{max}}}\leq\left|h_{rd}\right|^2\leq\frac{\mu\sigma^2_d|h_{rs}|^2\phi}
    {\tau-\sigma_s^2}\right]}\mathcal{P}_B^{-1} , &\sigma_s^2\leq\tau\leq\rho_1,\\
    0,  &\tau>\rho_1.
  \end{array}
\right.
\end{align}
Then, substituting $F_{|h_{rd}|^2}(x) = 1 - e^{-x}$ into the above equation ($h_{rs}$ is perfectly known by S and thus it is not a random variable here) we achieve the desired result in \eqref{PFA_Q}.

Considering the maximum power constraint at R under $\Halt$ (i.e., $P_r^1 + P_{\Delta} \leq P_r^{\mathrm{max}}$) and following \eqref{decisions}, \eqref{Pw}, and \eqref{Pr1_Q}, the miss detection rate under the condition $\mathbb{B}$ is given by
\begin{align}
&\beta =\mathcal{P}\left[\frac{\mu(Q+\sigma^2_d)|h_{rs}|^2\phi}{|h_{rd}|^2}+\frac{Q |h_{rs}|^2}{|h_{rd}|^2}+\sigma_s^2<\tau\big{|}\mathbb{B}\right] \notag \\
&=\left\{
\begin{array}{ll}
    0,  &\tau<\sigma_s^2, \\
    {\mathcal{P}\!\left[|h_{rd}|^2\!\geq\!\frac{\left(\phi\mu\left(\sigma^2_d+Q\right)+Q\right)|h_{rs}|^2}{\tau-\sigma_s^2}\right]}\mathcal{P}_B^{-1}, &\sigma_s^2\leq\tau\leq \rho_2,\\
    1,  &\tau>\rho_2.
\end{array} \label{PMD_Q2}
\right.
\end{align}
Then, substituting $F_{|h_{rd}|^2}(x) = 1 - e^{-x}$ into \eqref{PMD_Q2} we achieve the desired result in \eqref{PMD_Q}.
\end{IEEEproof}

We note that the false alarm and miss detection rates given in Theorem~\ref{theorem1} are functions of the threshold $\tau$ and we next examine how S sets the value of $\tau$ to minimize its detection error probability in the following subsection.

\subsection{Optimization of the Detection Threshold at Source}

In this subsection, we derive the optimal value of the detection threshold $\tau$ that minimizes the detection error probability $\xi$ for the rate-control transmission scheme.
\begin{theorem}\label{theorem2}
The optimal threshold that minimizes $\xi$ for the rate-control transmission scheme is given by
\begin{align} \label{tau_ast_Q}
\tau^{\ast}=\left\{
  \begin{array}{ll}
    \rho_1,  &\tau^{\ddag}\leq\sigma_s^2,\\
    \min\left\{\tau^{\ddag}, \rho_1\right\}, & \tau^{\ddag} > \sigma_s^2,
  \end{array}
\right.
\end{align}
where
\begin{align}
\tau^{\ddag}&\triangleq\frac{(\phi\mu+1)Q|h_{rs}|^2}{\ln\left(\frac{\omega_1}{1-\omega_1}\left(1+\frac{(\phi\mu+1)Q}{\phi\mu\sigma^2_d}\right)\right)}+\sigma^2_s, \label{tau_ddag} \\
\omega_1&\triangleq \frac{1}{2}\exp\left\{-\frac{(\mu+1)Q}{P_r^{\mathrm{max}}}\right\}. \label{omega1}
\end{align}
\end{theorem}

\begin{IEEEproof}
As discussed before, S will perform detection whenever condition $\mathbb{B}$ is met and R can transmit covert message when condition $\mathbb{C}$ is guaranteed. In our work, we assume that R will transmit a covert message with probability 50\% when $\mathbb{C}$ is true. As per \eqref{P_b} and \eqref{P_c_Q}, the probability $\mathcal{P}(\Halt)$ is given by
\begin{align}\label{P_H1_Q}
\mathcal{P}(\Halt) &=  \frac{1}{2}\mathcal{P}\left[\mathbb{C}\big{|}\mathbb{B}\right] =\omega_1.
\end{align}
Then, $\mathcal{P}(\Hnull)$ is given by
\begin{align}\label{P_H0_Q}
\mathcal{P}(\Hnull) &=  1-\mathcal{P}(\Halt) =1-\omega_1.
\end{align}

Since $\rho_2> \rho_1$ as given in Theorem~\ref{theorem1}, following \eqref{PFA_Q} and \eqref{PMD_Q}, we have the detection error probability at S as
\begin{align} \label{fixed rate xi}
\xi=\left\{
  \begin{array}{ll}
    1-\omega_1, & \tau \leq \sigma_s^2, \\
    1\!-\!\omega_1\!-\!\mathcal{P}_B^{-1}[(1\!-\!\omega_1)\kappa_1(\tau)-\\
    \omega_1\kappa_2(\tau)],  &\sigma_s^2 <\tau\leq\rho_1, \\
    \omega_1\mathcal{P}_B^{-1}\kappa_2(\tau),  &\rho_1\leq\tau<\rho_2, \\
    \omega_1,  &\tau\geq\rho_2.
  \end{array}
\right.
\end{align}

We first note that $\xi = 1-\omega_1$ or $\omega_1$ are the worst case for S and thus S does not set $\tau \leq \sigma_s^2$ or $\tau > \rho_2$. Following \eqref{fixed rate xi}, we derive the first derivative of $\xi$ with respect to $\tau$ when $\rho_1\leq\tau<\rho_2$ as
\begin{align}\label{partial_xi_add}
\frac{\partial(\xi)}{\partial\tau}=\frac{\omega_1\mathcal{P}_B^{-1}\left(\phi\mu\left(\sigma^2_d+Q\right)+Q\right)|h_{rs}|^2}{(\tau-\sigma_s^2)^2}\kappa_2(\tau)>0.
\end{align}
This demonstrates that $\xi$ is an increasing function of $\tau$ when $\rho_1\leq\tau<\rho_2$. Thus, S will set $\rho_1$ as the threshold to minimize $\xi$ if $\rho_1\leq\tau<\rho_2$.
We next derive the first derivative of $\xi$ with respect to $\tau$ for $\sigma_s^2 < \tau \leq \rho_1$ as
\begin{align} \label{partial_xi}
&\frac{\partial(\xi)}{\partial\tau}=\frac{\mathcal{P}_B^{-1}|h_{rs}|^2}{(\tau-\sigma_s^2)^2}\Big[\omega_1\left(\phi\mu\left(\sigma^2_d+ Q\right)+Q\right)\kappa_2(\tau)- \notag \\
&~~~~~~~~(1-\omega_1)\phi\mu\sigma^2_d\kappa_1(\tau)\Big] \notag \\
&=\frac{\omega_1\mathcal{P}_B^{-1}\left(\phi\mu\left(\sigma^2_d+Q\right)+Q\right)|h_{rs}|^2\kappa_2(\tau)}{(\tau-\sigma_s^2)^2} \times \notag \\
&~~\left\{1-\frac{(1-\omega_1)\phi\mu\sigma^2_d}{\omega_1\left(\phi\mu(\sigma^2_d+Q)+Q\right)}\mathrm{exp}\left\{\frac{(\phi\mu+1)Q|h_{rs}|^2}{\tau-\sigma_s^2}\right\}\right\}  .
\end{align}
We note that $\omega_1\mathcal{P}_B^{-1}\left(\phi\mu\left(\sigma^2_d+Q\right)+Q\right)|h_{rs}|^2\kappa_2(\tau)/(\tau-\sigma_s^2)^2 >0$ due to $\sigma_s^2 < \tau$ and $\kappa_2(\tau) >0$ as given in Theorem~\ref{theorem1}. As such, without the constraint $\tau \leq \rho_1$, the value of $\tau$ that ensures ${\partial(\xi)}/{\partial\tau}=0$ in \eqref{partial_xi} is given by $\tau^{\ddag}$.
We note that ${\partial(\xi)}/{\partial\tau}<0$, for $\tau<\tau^{\ddag}$, and ${\partial(\xi)}/{\partial\tau}>0$, for $\tau>\tau^{\ddag}$. This is due to the term $\mathrm{exp}\{{(\phi\mu+1)Q|h_{rs}|^2}/(\tau-\sigma_s^2)\}$ in \eqref{partial_xi} is monotonically decreasing with respect to $\tau$. This indicates that $\tau^{\ddag}$ minimizes $\xi$ without the constraint $\tau \leq \rho_1$. We also note that $\xi$ given in \eqref{fixed rate xi} is a not a continuous function of $\tau$ following Theorem~\ref{theorem1} when $\tau^{\ddag}\leq\sigma_s^2$. This is due to that $1-\omega_1-\mathcal{P}_B^{-1}[(1-\omega_1)\kappa_1(\tau)-\omega_1\kappa_2(\tau)]$ is monotonically increasing with respect to $\tau$ when $\tau^{\ddag}\leq\sigma_s^2$. We note that $\xi$ is also monotonically increasing with respect to $\tau$ for $\rho_1\leq\tau<\rho_2$, thus will lead to $\omega_1\geq 1-\omega_1$. As such, if $\tau^{\ddag}\leq\sigma_s^2$, the optimal threshold is $\tau^{\ast}= \rho_1.$.
If $\tau^{\ddag}>\sigma_s^2$, following \eqref{partial_xi_add} and noting $\xi$ is a continuous function of $\tau$, we can conclude that the optimal threshold is $\tau^{\ast}= \min\left\{\tau^{\ddag}, \rho_1\right\}.$
This completes the proof of Theorem~\ref{theorem2}.
\end{IEEEproof}

Following Theorem~\ref{theorem2}, we obtain the minimum detection error probability at S in the following corollary.

\begin{corollary}\label{corollary1}
The minimum value of $\xi$ at S is
\begin{align}
\xi^{\ast}=\left\{
  \begin{array}{ll}
    (1-\omega_1)\Bigg\{1-\exp\left(\frac{\mu\sigma^2_d}{P_r^{\mathrm{max}}}\right)\times \\
    \left(1-\frac{\phi\mu\sigma^2_d}{\phi\mu\sigma^2_d+(\phi\mu+1) Q}\right)\times \\
    \left(\frac{\omega_1}{1-\omega_1}\left(1+\frac{(\phi\mu+1)Q}{\phi\mu\sigma^2_d}\right)\right)^{-\frac{\phi\mu\sigma^2_d}{(\phi\mu +1)Q}}\Bigg\},&\tau^{\ast}=\tau^{\ddag},\\
    \omega_1\exp\left\{-\frac{(\phi\mu+1)Q}{\phi P_r^{\mathrm{max}}}\right\}, &\tau^{\ast}=\rho_1.
  \end{array} \label{optimal_xi_Q}
\right.
\end{align}
\end{corollary}

\begin{IEEEproof}
Substituting $\tau^{\ast}$ into \eqref{fixed rate xi}, we obtain the minimum value of $\xi$ as $\xi^{\ast} = 1-\omega_1-\mathcal{P}_B^{-1}[(1-\omega_1)\kappa_1(\tau)-\omega_1\kappa_2(\tau)]$, which completes the proof of Corollary~\ref{corollary1}.
\end{IEEEproof}

Based on Theorem~\ref{theorem1}, Theorem~\ref{theorem2},  and Corollary~\ref{corollary1}, we draw the following useful insights.
\begin{remark}\label{remark1_add}
We conclude that detection error probability $\xi^{\ast}$ tends to 0 when R's additional covert power Q approaches infinity. This follows from \eqref{tau_ast_Q} for $\tau^{\ast}=\rho_1$, since when $Q \rightarrow \infty$ we have $\tau^{\ddag} < \sigma_s^2$ as per \eqref{tau_ddag} and thus $\tau^{\ast}=\rho_1$.
\end{remark}
\begin{remark}\label{remark2}
When the maximum power constraint $P_r^{\mathrm{max}}$ approaches infinity, the minimum detection error probability $\xi^{\ast}$ approaches a fixed value given by
\begin{align} \label{xi_ast_Pmax_lim}
\lim_{P_b^{\mathrm{max}} \rightarrow \infty} {\xi^{\ast}}&=\frac{1}{2}\Bigg\{1-
    \underbrace{\left(1-\frac{\phi\mu\sigma^2_d}{\phi\mu\sigma^2_d+(\phi\mu+1) Q}\right)}_{f_1(Q)}\times\notag \\ &~~~\underbrace{\left(1+\frac{(\phi\mu+1)Q}{\phi\mu\sigma^2_d}\right)^{-\frac{\phi\mu\sigma^2_d}{(\phi\mu +1)Q}}}_{f_2(Q)}\Bigg\}.
\end{align}
The result in \eqref{xi_ast_Pmax_lim} follows from \eqref{tau_ast_Q} for $\tau^{\ast}=\tau^{\ddag}$, since when $P_r^{\mathrm{max}} \rightarrow \infty$ we have $\rho_1 \rightarrow \infty$ as per (24) and thus $\rho_1>\tau^{\ddag}$ (then $\tau^{\ast}=\tau^{\ddag}$).
Following \eqref{xi_ast_Pmax_lim}, we can conclude that $\xi^{\ast}$ monotonically decreases with $Q$ when $P_r^{\mathrm{max}} \rightarrow \infty$.
In order to prove this conclusion, we next prove that $f_2(Q)$ in \eqref{xi_ast_Pmax_lim} monotonically increases with $Q$, since $f_1(Q)$ in \eqref{xi_ast_Pmax_lim} is a monotonically increasing function of $Q$. Defining ${(\phi\mu +1)Q}/{\mu\sigma^2_d}=x$, following (35) we have $f_2(Q) = f_2(x)$, where
\begin{align}\label{f3x}
f_2(x) = (1+x)^{-1/x}.
\end{align}
In order to determine the monotonicity of $f_2(x)$ with respect to $x$, we derive its first derivative as
\begin{align}
\frac{\partial f_2(x)}{\partial x}=\exp\left\{-\frac{\ln(1+x)}{x}\right\}\frac{(1+x)\ln(1+x)-x}{x^2(1+x)}.
\end{align}
We note that whether ${\partial f_2(x)}/{\partial x} > 0$ or ${\partial f_2(x)}/{\partial x} < 0$ depends on $g(x) \triangleq (1+x)\ln(1+x)-x$. As such, we derive the first derivative of $g(x)$ with respect to $x$ as
\begin{align}
\frac{\partial g(x)}{\partial x}=\ln(1+x).
\end{align}
Noting that $x \geq 0$ and ${\partial g(x)}/{\partial x} \geq 0$, we conclude that $g(x)$ monotonically decreases with $x$. Then, we have $g(x)\geq g(0)=0$ and thus ${\partial f_2(x)}/{\partial x} \geq 0$. This leads to that $f_2(Q)$ monotonically increases with $Q$ and thus $\xi^{\ast}$ monotonically decreases with $Q$ for $\tau^{\ast} = \tau^{\ddag}$.
\\
When $P_r^{\mathrm{max}}\rightarrow \infty$, $\omega_1$ approaches 1/2 as per \eqref{omega1} and $\xi^{\ast} = \omega_1 - \epsilon$ can be written as
\begin{align} \label{covert_constraint_epsilon}
&\underbrace{\left(1-\frac{\phi\mu\sigma^2_d}{\phi\mu\sigma^2_d+(\phi\mu+1) Q}\right)}_{f_1(Q)} \underbrace{\left(1+\frac{(\phi\mu+1)Q}{\phi\mu\sigma^2_d}\right)^{-\frac{\phi\mu\sigma^2_d}{(\phi\mu +1)Q}}}_{f_2(Q)} \notag  \\
&=2\epsilon.
\end{align}
Defining $y=\phi\mu\sigma^2_d/{((\phi\mu +1)Q)}$ and following the expression of $f_2(Q)$ in \eqref{covert_constraint_epsilon}, we have
\begin{align}
\lim_{y \rightarrow 0} f_2(Q) =\lim_{y \rightarrow 0} \left(\frac{y}{y+1}\right)^{y} =0^{0}=1.
\end{align}
As per \eqref{covert_constraint_epsilon}, for $y \rightarrow 0$ the approximated close-form expression of $Q^{\epsilon}$ is given by
\begin{align}
Q^{\epsilon}=\frac{\phi\mu\sigma_d^2}{(\phi\mu+1)}\left(\frac{1}{1-2\epsilon}-1\right).
\end{align}
\end{remark}
\begin{remark}\label{remark3}
We have that the minimum detection error probability $\xi^{\ast}$ tends to 0 when the data transmission rate $R_{sd}$ approaches 0 or infinity. As $R_{sd} \rightarrow 0$, as per \eqref{mu} we have $\mu \rightarrow 0$ and thus $\tau^{\ddag} \rightarrow \sigma_s^2$ (then optimal threshold $\tau^{\ast}$ is equal to $\tau^{\ddag}$) following \eqref{tau_ddag}. Then, from \eqref{optimal_xi_Q} for $\tau^{\ast}=\tau^{\ddag}$ we can see that $\xi^{\ast} \rightarrow 0$ as $\mu \rightarrow 0$. As $R_{sd} \rightarrow \infty$, following \eqref{mu} again we note that $\mu$ will be negative and thus the transmission from S to D fails, which leads to $\xi^{\ast} \rightarrow 0$ as discussed in Section III-A. This result means that there exists an optimal value of $R_{sd}$ that maximizes $\xi^{\ast}$ and thus maximizes the effective covert rate for given other system parameters. We will numerically examine the impact of $R_{sd}$ on covert communications in Section V.
\end{remark}

\subsection{Optimization of Effective Covert Rate}

In this section, we examine the effective covert rate achieved in the considered system subject to a covert constraint.

\subsubsection{Effective Covert Rate}
From \eqref{gamma_c}, the SINR of $\mathbf{x}_c$ at D in the rate-control transmission scheme is given as
\begin{align}\label{gamma_c Q}
\gamma_{\Delta} &= \frac{P_{\Delta}|h_{rd}|^2}{P_r^1|h_{rd}|^2 G^2 \sigma^2_r +\sigma^2_d} \notag \\
&=\frac{Q}{\frac{\mu(Q+\sigma^2_d)}{\eta|h_{sr}|^2+1}+\sigma^2_d},
\end{align}
where $\eta\triangleq P_s/\sigma^2_r$. Then, the covert rate achieved by R is $R_{\Delta} = \log_2(1+\gamma_{\Delta})$. As such, we can see that the covert rate is fixed when Q is fixed as per \eqref{gamma_c Q}. We next derive the effective covert rate, i.e., the covert rate averaged over all realizations of $|h_{rd}|^2$, in the following theorem.

\begin{theorem}\label{theorem3}
The achievable effective covert rate $R_c$ by R in the rate-control transmission scheme is derived as a function of $Q$ given by
\begin{align} \label{fixed rate effective rate}
R_c &= R_{\Delta}\mathcal{P}_C =\log_2\left(1+\frac{Q}{\frac{\mu(Q+\sigma^2_d)}{\eta|h_{sr}|^2+1}+\sigma^2_d}\right)\times \notag \\
&~~~\exp\left\{-\frac{\mu\sigma^2_d+\mu Q+Q}{P_r^{\mathrm{max}}}\right\}.
\end{align}
\end{theorem}

Based on Theorem~\ref{theorem3}, we note that $R_c$ is not an increasing function of $Q$ and thus $R_{\Delta}$, since as $Q$ increases $R_{\Delta}$ increases as per \eqref{fixed rate effective rate} while $\mathcal{P}_C$ decreases following \eqref{P_c_Q}. This indicates that there may exists an optimal value of $Q$ that maximizes the effective covert rate, which motivates our following optimization of $Q$ in the considered system model.

\subsubsection{Maximization of $R_c$ with the Covert Constraint}

As per \eqref{P_H1_Q} and \eqref{P_H0_Q}, note that $\omega_1 \leq 1/2$, the covert constraint is given by
\begin{align}
\xi^{\ast} \geq \min\left\{1-\omega,\omega\right\} - \epsilon=\omega_1- \epsilon.
\end{align}

Following Theorem~\ref{theorem2}, the optimal value of $Q$ that maximizes $R_c$ subject to the covert constraint $\xi^{\ast} \geq \omega_1 - \epsilon$ can be obtained through numerical search, which is given by
\begin{align} \label{optimazation_problem_Q}
Q^{\ast} = &\argmax_{Q} R_c, \\ \nonumber
&\text {s.t.} ~~~~\xi^{\ast} \geq \omega_1- \epsilon.
\end{align}
We note that the optimization problem \eqref{optimazation_problem_Q} is of one dimension, which can be solved by efficient numerical search. The maximum value of $R_c$ is then achieved by substituting $Q^{\ast}$ into \eqref{fixed rate effective rate}, which is denoted by $R_c^{\ast}$.

\section{Power-Control Transmission Scheme}
In this section, we consider the power-control transmission scheme, in which R transmits a covert message to D with a constant transmit power if possible. Specifically, we first determine R's transmit power in $\Halt$ and then analyze the detection error probability at S, based on which we also derive S's optimal detection threshold. Furthermore, we derive the effective covert rate achieved by the power-control transmission scheme.
\subsection{Transmit Power at Relay}
Following \eqref{gamma_d2}, when $C_{sd} = R_{sd}$ we have
\begin{align}\label{Pr1}
P_r^{1}=\mu P_{\Delta}+\frac{\mu\sigma^2_d}{|h_{rd}|^2}.
\end{align}
We note that $C_{sd} = R_{sd}$ requires $C_{sd}^{*} \geq R_{sd}$ and thus $ |h_{rd}|^2 \geq {\mu\sigma^2_d}/{[P_r^{\mathrm{max}}-(\mu+1)P_{\Delta}]}$. Considering the maximum power constraint at R (i.e., $P_r^1 + P_{\Delta} \leq P_r^{\mathrm{max}}$ under this case), R has to give up the transmission of the covert message (i.e., $P_{\Delta} = 0$) when $P_r^1>P_r^{\mathrm{max}}-P_{\Delta}$ and sets $P_r^1$ the same as $P_r^0$ given in  \eqref{PR0}. This is due to the fact that S knows $h_{rs}$ and it can detect the covert transmission with probability one when the total transmit power of R is greater than $P_r^{\mathrm{max}}$. Then, the transmit power of $\mathbf{x}_r$ under $\Halt$ for the power-control transmission scheme is given by
\begin{align}
P_r^{1}=\left\{
  \begin{array}{ll}
    \mu P_{\Delta}+\frac{\mu\sigma^2_d}{|h_{rd}|^2}, & |h_{rd}|^2 \geq \frac{\mu\sigma^2_d}{P_r^{\mathrm{max}}-(\mu+1)P_{\Delta}}, \\
    \frac{\mu\sigma^2_d}{|h_{rd}|^2}, & \frac{\mu\sigma^2_d}{P_r^{\mathrm{max}}} \leq |h_{rd}|^2 < \frac{\mu\sigma^2_d}{P_r^{\mathrm{max}}-(\mu+1)P_{\Delta}}, \\
    0,  &|h_{rd}|^2 <\frac{\mu\sigma^2_d}{P_r^{\mathrm{max}}}.
  \end{array}
\right. \label{PR1}
\end{align}
As per \eqref{PR1}, we note that R also does not transmit a covert message when it cannot support the transmission from S to D (i.e., when $|h_{rd}|^2 <{\mu\sigma^2_d}/{P_r^{\mathrm{max}}}$). This is due to the fact that a transmission outage occurs when $|h_{rd}|^2 <{\mu\sigma^2_d}/{P_r^{\mathrm{max}}}$ and R or D would request a retransmission from S, which enables S to detect R's covert transmission with probability one if this covert transmission happened, since R cannot and thus does not forward S's information to D when $|h_{rd}|^2 <{\mu\sigma^2_d}/{P_r^{\mathrm{max}}}$. In summary, R could possibly transmit a covert message without being detected only when the condition $ |h_{rd}|^2 \geq {\mu\sigma^2_d}/{[P_r^{\mathrm{max}}-(\mu+1)P_{\Delta}]}$ is guaranteed. We again denote this necessary condition for covert communication as $\mathbb{C}$. Noting $F_{|h_{rd}|^2}(x) = 1 - e^{-x}$, the probability that $\mathbb{C}$ is guaranteed is given by
\begin{align}\label{Pc_delta}
\mathcal{P}_C = \exp\left\{-\frac{\mu\sigma^2_d}{P_r^{\mathrm{max}}-(\mu+1)P_{\Delta}}\right\}.
\end{align}
We note that $\mathcal{P}_C$ is a monotonically decreasing function of $P_{\Delta}$, which indicates that the probability that R can transmit a covert message (without being detected with probability one) decreases as $P_{\Delta}$ increases. Following \eqref{Pr1} and noting $P_r^1 + P_{\Delta} \leq P_r^{\mathrm{max}}$, we have $P_r^{\mathrm{max}} > (\mu+1)P_{\Delta}$.

\subsection{Detection Error Probability at Source}

In this subsection, we derive S's false alarm rate, i.e., $\alpha=\mathcal{P}(\Hoalt|\Hnull)$, and miss detection rate, i.e., $\beta=\mathcal{P}(\Honull|\Halt)$.

\begin{theorem}\label{theorem4}
When the condition $\mathbb{B}$ is guaranteed, for a given $\tau$, the false alarm and miss detection rates at S are derived as
\begin{align}
\alpha&=\left\{
  \begin{array}{ll}
    1,  &\tau<\sigma_s^2, \\
    1-\mathcal{P}_B^{-1}\kappa_1(\tau), & \sigma_s^2\leq\tau\leq\rho_1,\\
    0,  &\tau>\rho_1,
  \end{array}
\right. \label{PFA}\\
\beta&=\left\{
  \begin{array}{ll}
    0,  &\tau<\rho_3, \\
    \mathcal{P}_B^{-1}\kappa_3(\tau) , & \rho_3\leq\tau\leq \rho_4,\\
    1,  &\tau>\rho_4,
  \end{array}
\right. \label{PMD}
\end{align}
where
\begin{align}
&\rho_3\triangleq(\phi\mu+1)P_{\Delta}|h_{rs}|^2+\sigma_s^2, \notag \\
&\rho_4\triangleq \left(P_r^{\mathrm{max}}\phi+(\phi\mu+1)P_{\Delta}\right)|h_{rs}|^2+\sigma^2_s, \notag \\
&\kappa_3(\tau)\triangleq\mathrm{exp}\left\{-\frac{\phi\mu\sigma^2_d|h_{rs}|^2}{\tau-\rho_3}\right\}, \notag
\end{align}
and $\rho_1$ and $\kappa_1(\tau)$ are defined in \eqref{rho1_ex}.
\end{theorem}
\begin{IEEEproof}
Considering the maximum power constraint at R under $\Hnull$ (i.e., $P_r^0 \leq P_r^{\mathrm{max}}$) and following \eqref{PR0}, \eqref{decisions}, and \eqref{Pw}, the false alarm rate under the condition $\mathbb{B}$ is given by
\begin{align}
&\alpha=\mathcal{P}\left[\frac{\mu\sigma^2_d}{|h_{rd}|^2}|h_{rs}|^2\phi+\sigma_s^2\geq\tau\big{|}\mathbb{B}\right]  \notag\\
&=\left\{
  \begin{array}{ll}
    1,  &\tau<\sigma_s^2, \\
    {\mathcal{P}\!\left[\frac{\mu\sigma^2_d}{P_r^{\mathrm{max}}}\!\leq\!\left|h_{rd}\right|^2\!\leq\!\frac{\mu\sigma^2_d|h_{rs}|^2\phi}
    {\tau-\sigma_s^2}\right]}\mathcal{P}_B^{-1} , &\sigma_s^2\leq\tau\leq\rho_1,\\
    0,  &\tau>\rho_1.
  \end{array}
\right.
\end{align}
Then, substituting $F_{|h_{rd}|^2}(x) = 1 - e^{-x}$ into the above equation we achieve the desired result in \eqref{PFA}.

We first clarify that we have $\rho_3 < \rho_4$. Then, considering the maximum power constraint at R under $\Halt$ (i.e., $P_r^1 + P_{\Delta} \leq P_r^{\mathrm{max}}$) and following, \eqref{decisions}, \eqref{Pw}, and \eqref{PR1}, the miss detection rate under the condition $\mathbb{B}$ is given by
\begin{align}
&\beta=\mathcal{P}\left[\left(\mu P_{\Delta}+\frac{\mu\sigma^2_d}{|h_{rd}|^2}\right)|h_{rs}|^2\phi+P_{\Delta}|h_{rs}|^2+\sigma_s^2<\tau\big{|}\mathbb{B}\right] \notag \\
&=\left\{
\begin{array}{ll}
    0,  &\tau<\rho_3, \\
    {\mathcal{P}\!\left[|h_{rd}|^2\!\geq\!\frac{\phi\mu\sigma^2_d|h_{rs}|^2}{\tau-(\phi\mu+1)P_{\Delta}|h_{rs}|^2-\sigma_s^2}\right]}\!\mathcal{P}_B^{-1}, &\rho_3\leq\tau\leq \rho_4,\\
    1,  &\tau>\rho_4.
\end{array}
\right.
\end{align}
Then, substituting $F_{|h_{rd}|^2}(x) = 1 - e^{-x}$ into the above equation we achieve the desired result in \eqref{PMD}.
\end{IEEEproof}

We note that the false alarm and miss detection rates given in Theorem~\ref{theorem4} are functions of the threshold $\tau$ and we examine how S sets the value of $\tau$ to minimize its detection error probability in the following subsection.

\subsection{Optimization of the Detection Threshold at Source}
In this subsection, we first derive a constraint (i.e., an upper bound) on $P_{\Delta}$ to ensure a non-zero detection error probability at S. Then, under this constraint we derive the lower and upper bounds on the optimal value of $\tau$ that minimizes the detection error probability $\xi$ for the power-control transmission scheme.

\begin{theorem}\label{theorem5}
R's transmit power of the covert message $P_{\Delta}$ should satisfy
\begin{align}\label{P_condition}
P_{\Delta}\leq P_{\Delta}^u \triangleq \frac{\phi P_r^{\mathrm{max}}}{\phi\mu+1}
\end{align}
in order to guarantee $\xi > 0$ and when \eqref{P_condition} is guaranteed the optimal $\tau$ at S that minimizes $\xi$ should satisfy $\rho_3 \leq \tau^{\ast} \leq \rho_1$.
\end{theorem}
\begin{IEEEproof}
As discussed before, S will perform detection whenever condition $\mathbb{B}$ is met. In our work, we assume that R will transmit a covert message with probability 50\% when $\mathbb{C}$ is guaranteed. As per \eqref{P_b} and \eqref{Pc_delta}, the probability $\mathcal{P}(\Halt)$ is given by
\begin{align}\label{P_H1_pdelta}
\mathcal{P}(\Halt) &=  \frac{1}{2}\mathcal{P}(\mathbb{C}\big{|}\mathbb{B})=\omega_2,
\end{align}
where
\begin{align} \omega_2\triangleq\frac{1}{2}\exp\left\{-\frac{\mu(\mu+1)\sigma_d^2P_{\Delta}}{(P_r^{\mathrm{max}}\left(P_r^{\mathrm{max}}-(\mu+1)P_{\Delta}\right))}\right\}.
\end{align}

Then, $\mathcal{P}(\Hnull)$ is given by
\begin{align}\label{P_H0_pdelta}
\mathcal{P}(\Hnull) =  1-\mathcal{P}(\Halt)=1-\omega_2.
\end{align}

When $\rho_1< \rho_3$ that requires $P_{\Delta}> \phi P_r^{\mathrm{max}}/(\phi\mu+1)$ as per Theorem~\ref{theorem4}, following \eqref{PFA} and \eqref{PMD}, we have
\begin{align} \label{fixed power xi case 1}
\xi=\left\{
  \begin{array}{ll}
    1-\omega_2, & \tau \leq \sigma_s^2, \\
    (1-\omega_2)\left(1-\mathcal{P}_B^{-1}\kappa_1(\tau)\right),  & \sigma_s^2<\tau < \rho_1, \\
    0,  &\rho_1 \leq  \tau\leq\rho_3, \\
    \omega_2\mathcal{P}_B^{-1}\kappa_3(\tau),  &\rho_3<\tau<\rho_4, \\
    \omega_2,  &\tau\geq \rho_4.
  \end{array}
\right.
\end{align}
This indicates that S can simply set $\tau\in\left[\rho_1, \rho_3 \right]$ to ensure $\xi=0$ when $P_{\Delta}> \phi P_r^{\mathrm{max}}/(\phi\mu+1)$, i.e.,
S can detect the covert transmission with probability one. As such, $P_{\Delta}$ should satisfy \eqref{P_condition} in order to guarantee $\xi >0$.

We next prove $\rho_3 \leq \tau^{\ast} \leq \rho_1$.
When $P_{\Delta}\leq \phi P_r^{\mathrm{max}}/(\phi\mu+1)$, i.e., $\rho_3< \rho_1$, following \eqref{PFA} and \eqref{PMD}, we have
\begin{align} \label{fixed power xi case 2}
\xi=\left\{
  \begin{array}{ll}
    1-\omega_2, & \tau \leq \sigma_s^2, \\
    (1-\omega_2)\left(1-\mathcal{P}_B^{-1}\kappa_1(\tau)\right),  & \sigma_s^2<\tau \leq \rho_3, \\
    1-\omega_2-\mathcal{P}_B^{-1}\times \\
    \left[(1-\omega_2)\kappa_1(\tau)-\omega_2\kappa_3(\tau)\right],  &\rho_3<\tau<\rho_1, \\
    \omega_2\mathcal{P}_B^{-1}\kappa_3(\tau),  &\rho_1\leq\tau<\rho_4, \\
    \omega_2,  &\tau\geq\rho_4.
  \end{array}
\right.
\end{align}
Obviously, S will not set $\tau \leq \sigma_s^2$ or $\tau\geq\rho_4$, since $\xi = 1-\omega_1$ or $\omega_1$ are the worst case for S.

For $\sigma_s^2<\tau \leq \rho_3$, we derive the first derivative of $\xi$ with respect to $\tau$ as
\begin{align} \label{partial_xi_cond1}
\frac{\partial(\xi)}{\partial\tau}=-\frac{(1-\omega_2)\mathcal{P}_B^{-1}\phi\mu\sigma^2_d|h_{rs}|^2}{(\tau-\sigma_s^2)^2}\kappa_1(\tau)<0.
\end{align}
This demonstrates that $\xi$ is a decreasing function of $\tau$ when $\sigma_s^2<\tau \leq \rho_3$.
For $\rho_1\leq\tau<\rho_4$, we derive the first derivative of $\xi$ with respect to $\tau$ as
\begin{align} \label{partial_xi_cond2}
\frac{\partial(\xi)}{\partial\tau}=\frac{\omega_2\mathcal{P}_B^{-1}\phi\mu\sigma^2_d|h_{rs}|^2}{(\tau-\rho_3)^2}\kappa_3(\tau)>0.
\end{align}
This proves that $\xi$ is an increasing function of $\tau$ when $\rho_1\leq\tau<\rho_4$.
Noting that $\xi$ is a continuous function of $\tau$ and considering \eqref{partial_xi_cond1} and \eqref{partial_xi_cond2}, we can conclude that $\tau^{\ast}$ should satisfy $\rho_3 \leq \tau^{\ast} \leq \rho_1$, no mater what is the value of $\xi$ for $\rho_3<\tau<\rho_1$.
\end{IEEEproof}

The lower and upper bounds on $\tau^{\ast}$ given in Theorem~\ref{theorem5} significantly facilitate the numerical search for $\tau^{\ast}$ at S. Then, following Theorem~\ref{theorem5} and \eqref{fixed power xi case 2}, $\tau^{\ast}$ can be obtained through
\begin{align}\label{tau_optimal fix power}
\tau^{\ast} = \argmin_{\rho_3 \leq \tau \leq \rho_1} \left\{1-\omega_2-\mathcal{P}_B^{-1}\left[(1-\omega_2)\kappa_1(\tau)-\omega_2\kappa_3(\tau)\right]\right\}.
\end{align}
Substituting $\tau^{\ast}$ into \eqref{fixed power xi case 2}, we can obtain the minimum detection error probability $\xi^{\ast}$ for the power-control transmission scheme.

\subsection{Optimization of Effective Covert Rate}

In this section, we examine the effective covert rate achieved by the power-control transmission scheme subject to the covert constraint.

\subsubsection{Effective Covert Rate}
Following \eqref{gamma_c}, the SINR at destination for covert communication is given as
\begin{align}\label{gammac_fix_p}
\gamma_{\Delta} &= \frac{P_{\Delta}|h_{rd}|^2}{P_r^1|h_{rd}|^2 G^2 \sigma^2_r +\sigma^2_d} \notag \\
&=\frac{P_{\Delta}(\eta|h_{sr}|^2+1)|h_{rd}|^2}{\mu P_{\Delta}|h_{rd}|^2+(\eta|h_{sr}|^2+\mu+1)\sigma^2_d}.
\end{align}
Then, the covert rate achieved by R is $R_{\Delta} = \log_2(1+\gamma_{\Delta})$. We next derive the effective covert rate, i.e., averaged $R_{\Delta}$ over all realizations of $|h_{rd}|^2$, in the following theorem.

\begin{theorem}\label{theorem6}
The achievable effective covert rate $R_c$ by R with the power-control transmission scheme is derived as a function of $P_{\Delta}$ given by
\begin{align}\label{fixed power effective rate}
R_c&=\frac{1}{\ln 2}\exp\left\{-\frac{\mu \sigma^2_d}{P_r^{\mathrm{max}}-(\mu+1)P_{\Delta}}\right\}\times \notag \\
&\left[\ln\left(\frac{\beta_1}{\beta_2}\right)+e^{\frac{\beta_2}{\alpha_2}}\mathbf{Ei}\left(-\frac{\beta_2}{\alpha_2}\right)- e^{\frac{\beta_1}{\alpha_1}}\mathbf{Ei}\left(-\frac{\beta_1}{\alpha_1}\right)\right],
\end{align}
where
\begin{align}
\beta_1&\triangleq[\eta|h_{sr}|^2+\mu+1](P_r^{\mathrm{max}}-P_{\Delta})\sigma^2_d, \notag \\
\beta_2&\triangleq\left\{\frac{\eta|h_{sr}|^2+\mu+1}{[P_r^{\mathrm{max}}-(\mu+1)P_{\Delta}]^{-1}}+\mu^2P_{\Delta}\right\}\sigma^2_d, \notag \\
\alpha_1&\triangleq P_{\Delta}\left[\eta|h_{sr}|^2+(\mu+1)\right]\left[P_r^{\mathrm{max}}-(\mu+1)P_{\Delta}\right], \notag \\
\alpha_2&\triangleq\mu P_{\Delta}[P_r^{\mathrm{max}}-(\mu+1)P_{\Delta}], \notag
\end{align}
and the exponential integral function $\mathbf{Ei}(\cdot)$ is given by
\begin{align}
\mathbf{Ei}(x)=-\int_{-x}^{\infty} \frac{e^{-t}}{t} dt,~~~[x<0].
\end{align}
\end{theorem}

\begin{IEEEproof}
A positive covert rate is only achievable under the condition $\mathbb{C}$ and thus $R_c$ is given by
\begin{align}
R_c &=\int^{\infty}_{\frac{\mu \sigma^2_d}{P_r^{\mathrm{max}}-(\mu+1)P_{\Delta}}} R_{\Delta} f(|h_{rd}|^2) d |h_{rd}|^2 \notag \\
&\overset{a}{=}\frac{1}{\ln 2}\exp\left\{-\frac{\mu \sigma^2_d}{P_r^{\mathrm{max}}-(\mu+1)P_{\Delta}}\right\} \times \notag \\ &\int^{\infty}_{0}\ln\left(\frac{\beta_1+\alpha_1 x}{\beta_2+\alpha_2 x}\right)e^{-x} dx,\label{R_C_details}
\end{align}
where $\overset{a}{=}$ is achieved by setting $x=|h_{rd}|^2-{\mu \sigma^2_d}/[{P_r^{\mathrm{max}}-(\mu+1)P_{\Delta}}]$.
We then solve the integral in \eqref{R_C_details} with the aid of \cite[Eq. (4.337.1)]{Gradshteyn2007table}
\begin{align}
\int^{\infty}_{0} e^{-\nu x}\ln(\theta+x) dx=\frac{1}{\nu}\left[\ln \theta+e^{\nu \theta }\mathbf{Ei}(-\theta \nu)\right],
\end{align}
and achieve the result given in \eqref{fixed power effective rate}.
\end{IEEEproof}

Based on Theorem~\ref{theorem6}, we note that $R_c$ is not an increasing function of $P_{\Delta}$, since as $P_{\Delta}$ increases $R_{\Delta}$ increases but $\mathcal{P}_C$ (i.e., the probability that the condition $\mathbb{C}$ is guaranteed) decreases. This motivates our following optimization of $P_{\Delta}$ in order to maximize the effective covert rate subject to the covert constraint.

\subsubsection{Maximization of $R_c$ with the Covert Constraint}
As per \eqref{P_H1_pdelta} and \eqref{P_H0_pdelta}, note that $\omega_2 \leq 1/2$, the covert constraint is given by
\begin{align}
\xi^{\ast} \geq \min\left\{1-\omega,\omega\right\} - \epsilon=\omega_2- \epsilon.
\end{align}

Following Theorem~\ref{theorem5} the optimal value of $P_{\Delta}$ that maximizes $R_c$ subject to this constraint can be obtained through
\begin{align} \label{optimazation_problem_1}
P_{\Delta}^{\ast} = &\argmax_{0 \leq P_{\Delta} \leq P_{\Delta}^u} R_c \\ \nonumber
&\text {s.t.} ~~~~\xi^{\ast} \geq \omega_2- \epsilon.
\end{align}

We note that this is a two-dimensional optimization problem that can be solved by efficient numerical searches. Specifically, for each given $P_{\Delta}$, $\xi^{\ast}$ should be obtained based on \eqref{tau_optimal fix power} where $\tau^{\ast}$ is also numerically searched. We note that the numerical search of $P_{\Delta}^{\ast}$ and $\tau^{\ast}$ is efficient since their lower and upper bounds are explicitly given. The maximum value of $R_c$ is denoted by $R_c^{\ast}$.

\section{Numerical Results}

In this section, we first present numerical results to verify our analysis on the performance of covert communications in relay networks. Then, we provide a thorough performance comparison between the rate-control and power-control transmission schemes. Based on our examination, we draw many useful insights with regard to the impact of some system parameters (e.g., $P_r^{\mathrm{max}}$, $R_{sd}$, and $\epsilon$ ) on covert communications in wireless relay networks.

\subsection{Rate-Control Transmission Scheme}

\begin{figure}[!t]
    \begin{center}
    \includegraphics[width=3.5in, height=2.9in]{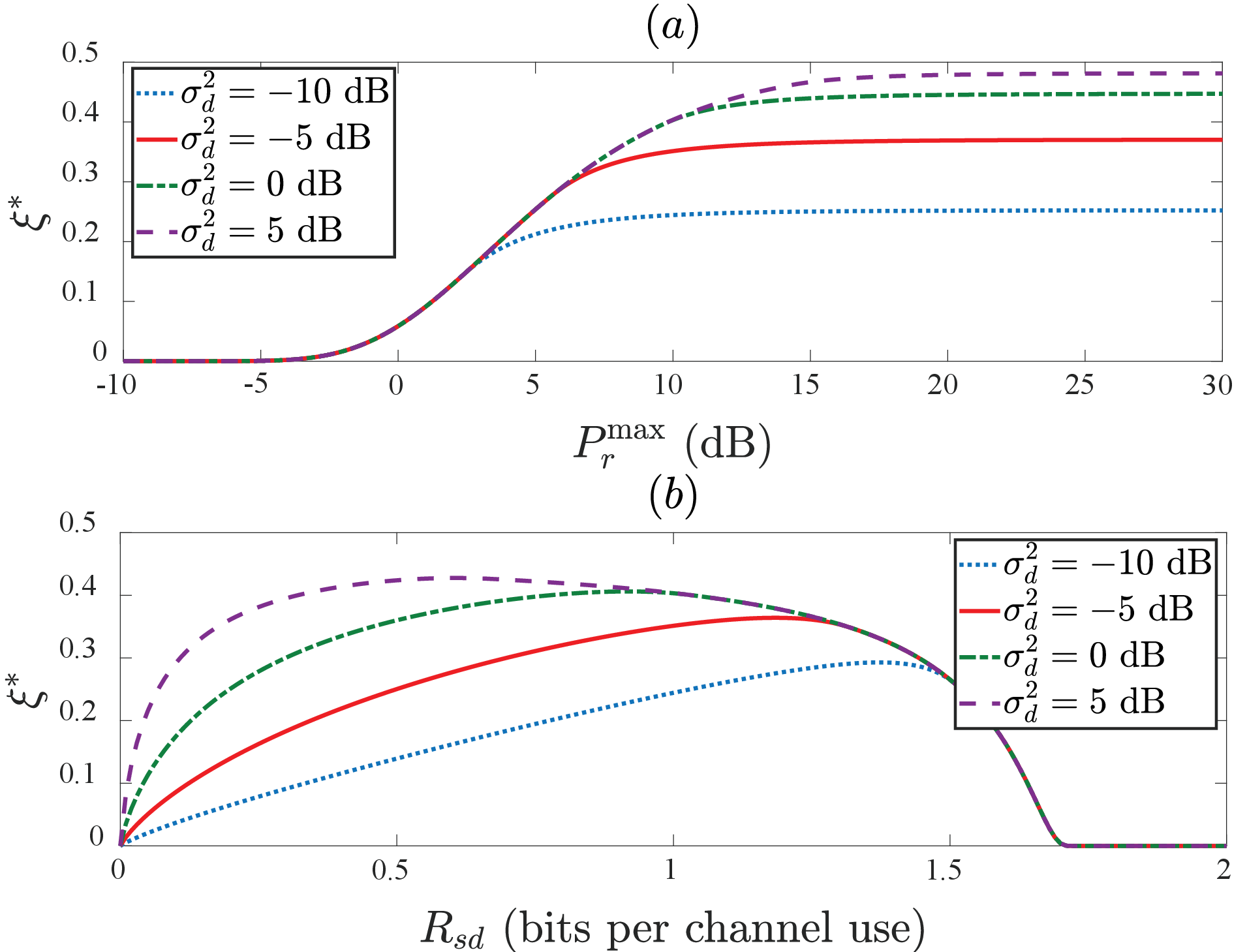}
    \caption{(a) $\xi^{\ast}$ versus $P_r^{\mathrm{max}}$ with different value of $\sigma^2_d$ for the rate-control transmission scheme, where $P_{s}=10$~dB, $\sigma^2_r=0$~dB, $R_{sd}=1~\mathrm{bits~per~channel~use}$, $|h_{sr}|^2=|h_{rs}|^2=1$, and $Q=0.1$. (b) $\xi^{\ast}$ versus $R_{sd}$ with different value of $\sigma^2_d$ for the rate-control transmission scheme, where $P_{s}=P_r^{\mathrm{max}}=10$~dB, $\sigma^2_r=0$~dB, $|h_{sr}|^2=|h_{rs}|^2=1$, and $Q=0.1$.}\label{fig2}
    \end{center}
\end{figure}

In Fig.~\ref{fig2}~(a), we plot the minimum detection error probability $\xi^{\ast}$ versus R's maximum transmit power $P_r^{\mathrm{max}}$ and observe that $\xi^{\ast}$ increases with $P_r^{\mathrm{max}}$. This shows that the covert transmission becomes easier as the desired performance of the normal transmission increases, since the transmission outage probability decreases with $P_r^{\mathrm{max}}$ for a fixed $R_{sd}$. We also observe $\xi^{\ast}$ approach to a specific value as  $P_r^{\mathrm{max}} \rightarrow \infty$, which is discussed in Remark~\ref{remark2}. This observation demonstrates that the covert transmission can still be possibly detected by S even without the maximum power constraint at R.
In Fig.~\ref{fig2} (b), we plot $\xi^{\ast}$ versus the transmission rate from S to D (i.e., $R_{sd}$). We first observe that $\xi^{\ast}$ is not a monotonic function of $R_{sd}$ and $\xi^{\ast} \rightarrow 0$ as $R_{sd} \rightarrow 0$ or $R_{sd} \rightarrow \infty$. This observation indicates that there may exist an optimal value of $R_{sd}$ that maximizes $\xi^{\ast}$. In Fig.~\ref{fig2}, we finally observe that $\xi^{\ast}$ is a monotonic increasing function of $\sigma_d^2$.

\begin{figure}[!t]
    \begin{center}
    \includegraphics[width=3.5in, height=2.9in]{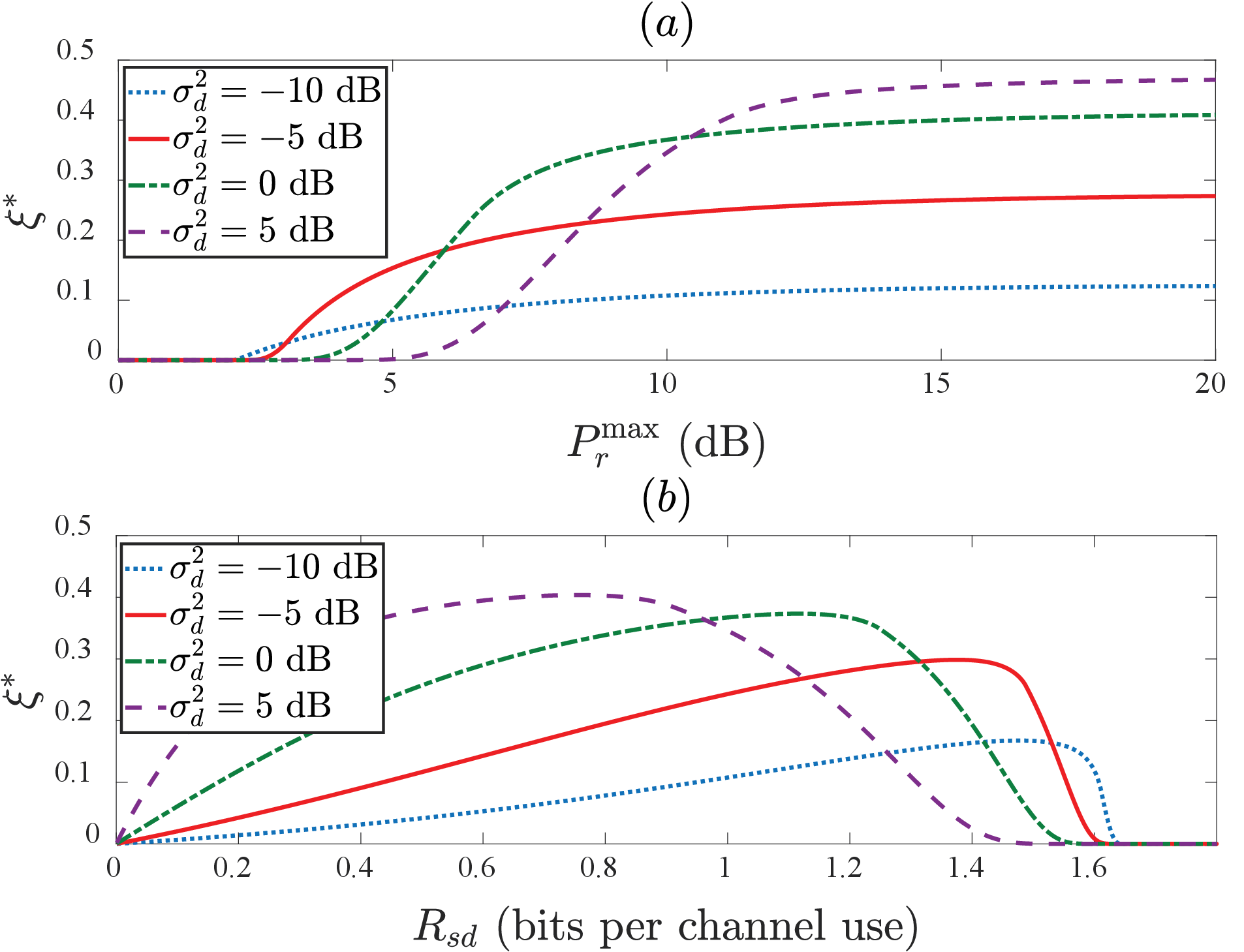}
    \caption{(a) $\xi^{\ast}$ versus $P_r^{\mathrm{max}}$ with different value of $\sigma^2_d$ for the power-control transmission scheme, where $P_{s}=10$~dB, $\sigma^2_r=0$~dB, $R_{sd}=1~\mathrm{bits~per~channel~use}$, $|h_{sr}|^2=|h_{rs}|^2=1$, and $P_{\Delta}=-10$~dB. (b) $\xi^{\ast}$ versus $R_{sd}$ with different value of $\sigma^2_d$ for the power-control transmission scheme, where $P_{s}=P_r^{\mathrm{max}}=10$~dB, $\sigma^2_r=0$~dB, $|h_{sr}|^2=|h_{rs}|^2=1$, and $P_{\Delta}=-10$~dB.}\label{fig3}
    \end{center}
\end{figure}

\subsection{Power-Control Transmission Scheme}

In Fig.~\ref{fig3}~(a), we plot the minimum detection error probability $\xi^{\ast}$ versus R's maximum transmit power $P_r^{\mathrm{max}}$ and observe that $\xi^{\ast}$ increases with $P_r^{\mathrm{max}}$. This shows that the covert transmission becomes easier as the desired performance of the normal transmission increases, since the transmission outage probability decreases with $P_r^{\mathrm{max}}$ for a fixed $R_{sd}$. We also observe $\xi^{\ast}$ does not approach 1/2 (but a specific value that is lower than 1/2) as  $P_r^{\mathrm{max}} \rightarrow \infty$, which is the same as the result discussed in Remark~\ref{remark2} for the rate-control transmission scheme. This observation demonstrates that the covert transmission can still be possibly detected by S even without the maximum power constraint at R. Fig.~\ref{fig3} (b), we plot $\xi^{\ast}$ versus the transmission rate from S to D (i.e., $R_{sd}$). We first observe that $\xi^{\ast}$ is not a monotonic function of $R_{sd}$ and $\xi^{\ast} \rightarrow 0$ as $R_{sd} \rightarrow 0$ or $R_{sd} \rightarrow \infty$. This observation indicates that there may exist an optimal value of $R_{sd}$ that maximizes $\xi^{\ast}$. In Fig.~\ref{fig3}, we finally observe that $\xi^{\ast}$ is not a monotonic function of $\sigma_d^2$.

\subsection{Performance Comparisons between the Rate-Control and Power-Control Transmission Schemes}

\begin{figure}[!t]
    \begin{center}
    \includegraphics[width=3.5in, height=2.9in]{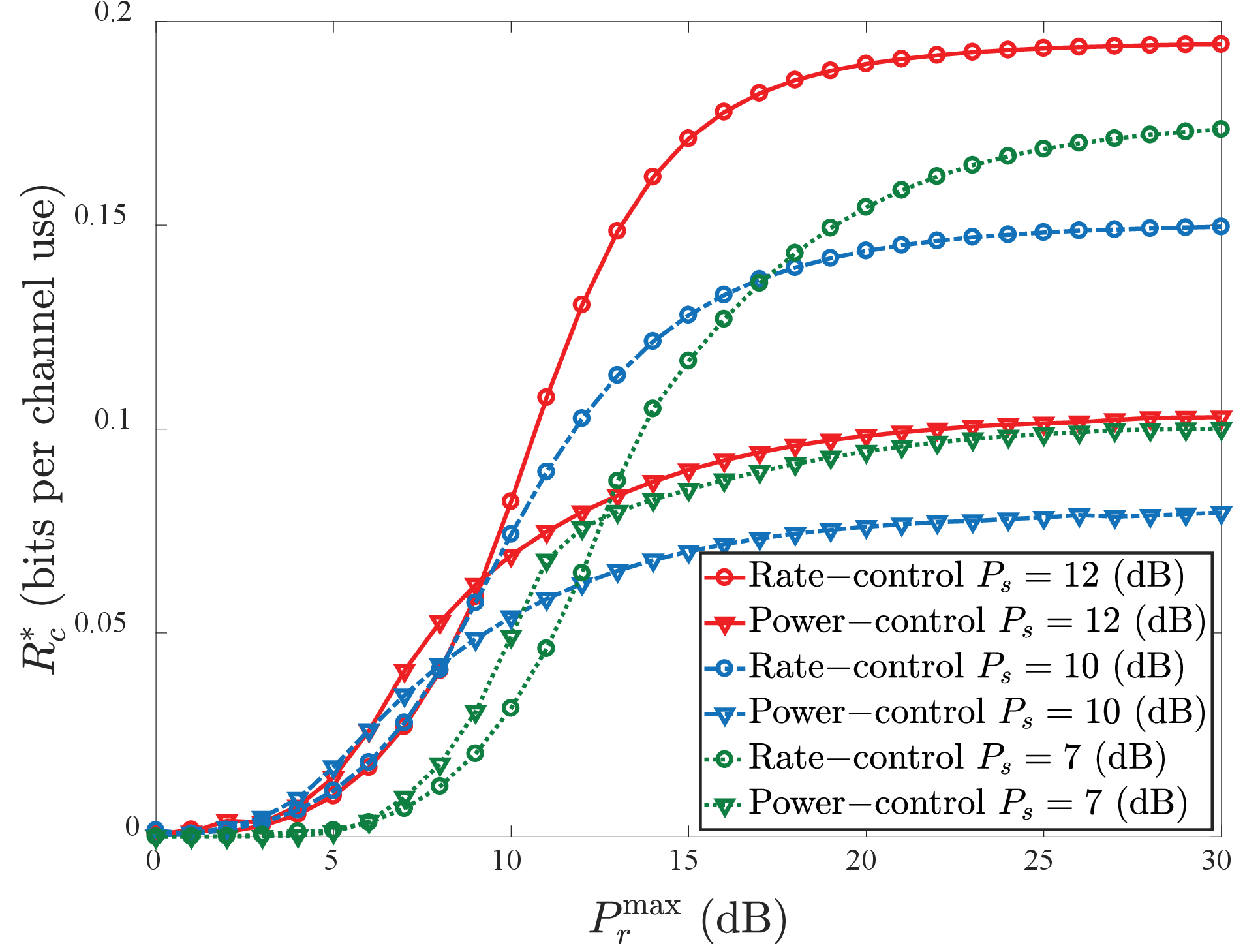}
    \caption{$R_c^{\ast}$ versus $P_r^{\mathrm{max}}$ under different value of $P_s$, where $\sigma^2_r=\sigma^2_d=0$~dB, $\epsilon = 0.1$, $R_{sd}=1~\mathrm{bits~per~channel~use}$, and $|h_{sr}|^2=1$.}\label{fig4}
    \end{center}
\end{figure}

\begin{figure}[!t]
    \begin{center}
    \includegraphics[width=3.5in, height=2.9in]{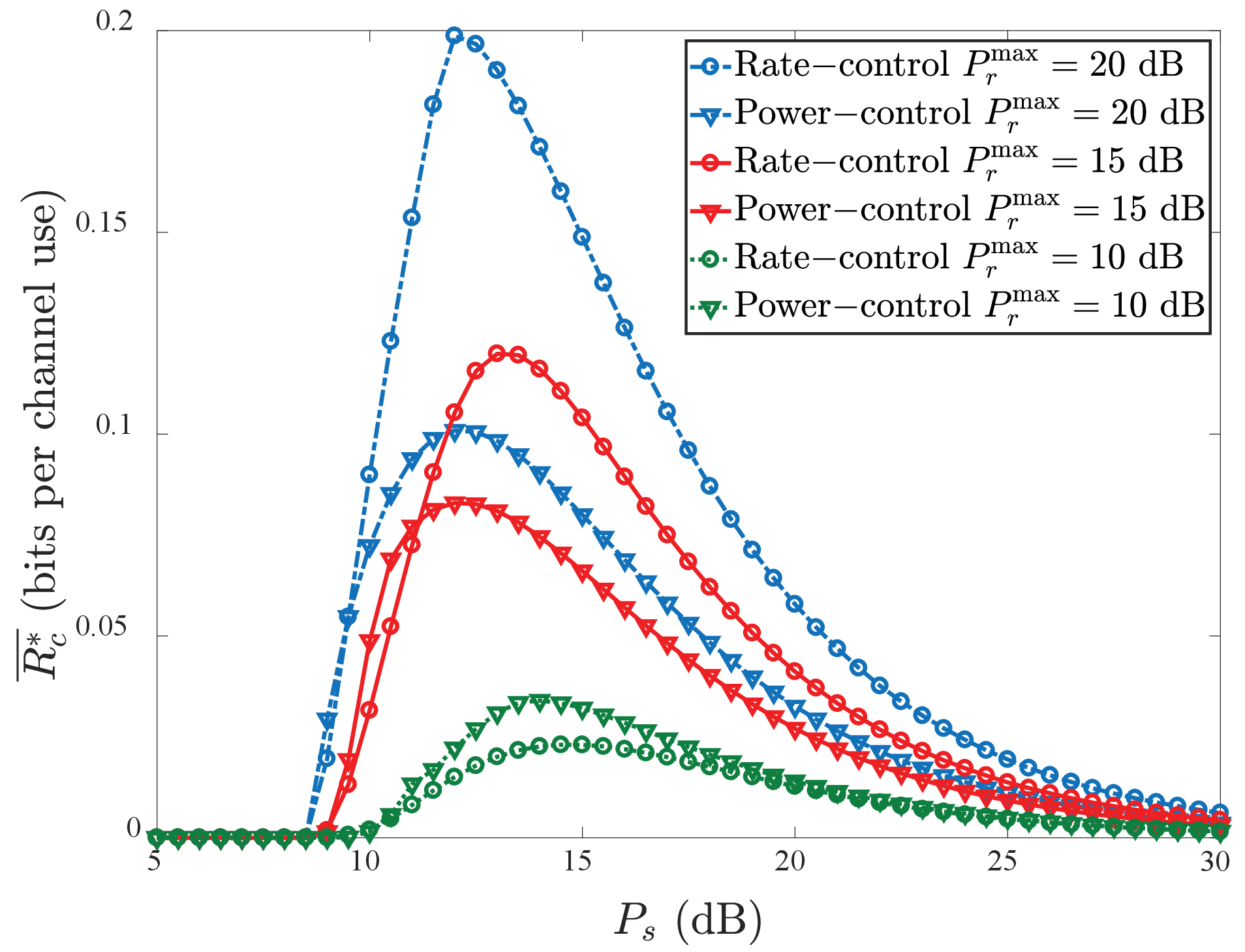}
    \caption{$\overline{R_c^{\ast}}$ versus $P_s$ under different value of $P_r^{\mathrm{max}}$, where $\sigma^2_r=\sigma^2_d=0$~dB, $\epsilon = 0.1$, and $R_{sd}=1.5~\mathrm{bits~per~channel~use}$.}\label{fig5}
    \end{center}
\end{figure}

Fig.~\ref{fig4} illustrates $R_c^{\ast}$ versus $P_r^{\mathrm{max}}$ with different values of $P_s$ for the rate-control and power-control transmission schemes using \eqref{optimazation_problem_Q} and \eqref{optimazation_problem_1}, respectively. In this figure, we first observe that for both schemes $R_c^{\ast}$ monotonically increases as $P_r^{\mathrm{max}}$ increases, which demonstrates that the covert message becomes easier to be transmitted when more power is available at R. we also observe that $R_c^{\ast}$ is not a monotonic function of $P_s$.
In Fig.~\ref{fig4}, it illustrates that the power-control transmission scheme outperforms the rate-control transmission scheme when  $P_r^{\mathrm{max}}$ is in the low regime. However, when $P_r^{\mathrm{max}}$ is larger than some specific values (e.g., when $P_r^{\mathrm{max}}\geq 13$~dB), the performance of rate-control transmission scheme is better than that of the power-control transmission scheme. This is mainly due to the fact that the transmit power constraints are not limits of the covert transmission when $P_r^{\mathrm{max}}$ is large, and thus under this case selecting a proper covert transmission rate (in the rate-control transmission scheme) can gain more benefit. We note that this observation demonstrates the significance of our work, since with our analysis R can easily determine which transmission is better under the specific system settings.

In Fig.~\ref{fig5}, we plot the averaged maximum effective covert rate, i.e., $\overline{R_c^{\ast}}$, which is achieved by averaging ${R}_c^{\ast}$ over $|h_{sr}|^2$, versus S's transmit power $P_s$ with different values of R's maximum transmit power $P_r^{\mathrm{max}}$. In this figure, we first observe that $\overline{R_c^{\ast}}$ is zero when $P_s$ is effectively small (e.g., due to the fact that S is far from R and D). This is due to the fact that when $P_s$ is sufficient small, the normal transmission from S to D with the fixed rate $R_{sd}$ may not be supported and R does not forward S's information to D. Meanwhile, the covert transmission from R to D cannot be achieved due to the lack of the shield from the normal transmission. We also observe that $\overline{R_c^{\ast}} \rightarrow 0$ when $P_s \rightarrow \infty$. This is due to the fact that $\phi$ given in \eqref{T} decreases (and thus $P_r^0 |h_{rs}|^2\phi$ and $P_r^1 |h_{rs}|^2\phi$ decrease) with $P_s$, which leads to a lower detection error probability at S as per \eqref{optimal_xi_Q} and \eqref{tau_optimal fix power} (i.e., it becomes easier for S to detect the covert transmission).
In Fig.~\ref{fig5}, we further observe that the achieved $\overline{R_c^{\ast}}$ decreases significantly as~$P_r^{\mathrm{max}}$ decreases (e.g., when R is with less transmit power than S), which demonstrates that it is the power constraint at R that mainly limits the performance of the covert transmission. Based on this observation, we can predict that $\overline{R_c^{\ast}} \rightarrow 0$ when $P_r^{\mathrm{max}} \rightarrow 0$. This is due to the fact that as~$P_r^{\mathrm{max}} \rightarrow 0$ R cannot support the normal transmission from S to D, not to mention the covert transmission from itself to D (due to the lack of the shield). Finally, we observe that the power-control transmission scheme outperforms the rate-control transmission scheme when $P_s$ or $P_r^{\mathrm{max}}$ is low. This observation is consistent with that found in Fig.~\ref{fig4}.

\section{Conclusion}

This work examined covert communication in one-way relay networks over quasi-static Rayleigh fading channels, in which R opportunistically transmits its own information to the destination covertly on top of forwarding S's message in AF mode, while S tries to detect this covert transmission. Specifically, we proposed the rate-control and power-control transmission schemes for R to convey covert information to D. We analyzed S's detection limits of the covert transmission from R to D in terms of the detection error probability and determined the achievable effective covert rates subject to $\xi^{\ast} \geq \min\{1-\omega, \omega\}-\epsilon$ for these two schemes. Our examination showed that the rate-control transmission scheme outperforms the power-control transmission scheme under some specific conditions, and otherwise the power-control transmission scheme outperforms the rate-control transmission scheme. As such, our conducted analysis enabled R to switch between these two strategies to achieve the maximum covert rate. Our investigation also demonstrated that covert communication in the considered relay networks is feasible and the effective covert rate achieved by R increases with its forwarding ability.

\bibliographystyle{IEEEtran}
\bibliography{IEEEfull,CC}

\begin{IEEEbiography}[{\includegraphics[width=1in,height=1.25in,clip,keepaspectratio]{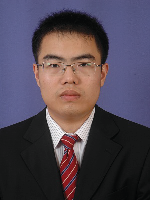}}]{Jinsong Hu} received the B.S. degree from the Nanjing University of Science and Technology, Nanjing, China, in 2013. He is currently pursuing the Ph.D. degree with the School of Electronic and Optical Engineering, Nanjing University of Science and Technology, Nanjing, China. He is also a Visiting Ph.D. Student with the Australian National University from 2017 to 2018. His research interests include array signal processing, covert communications, and physical layer security.
\end{IEEEbiography}
\begin{IEEEbiography}[{\includegraphics[width=1in,height=1.25in,clip,keepaspectratio]{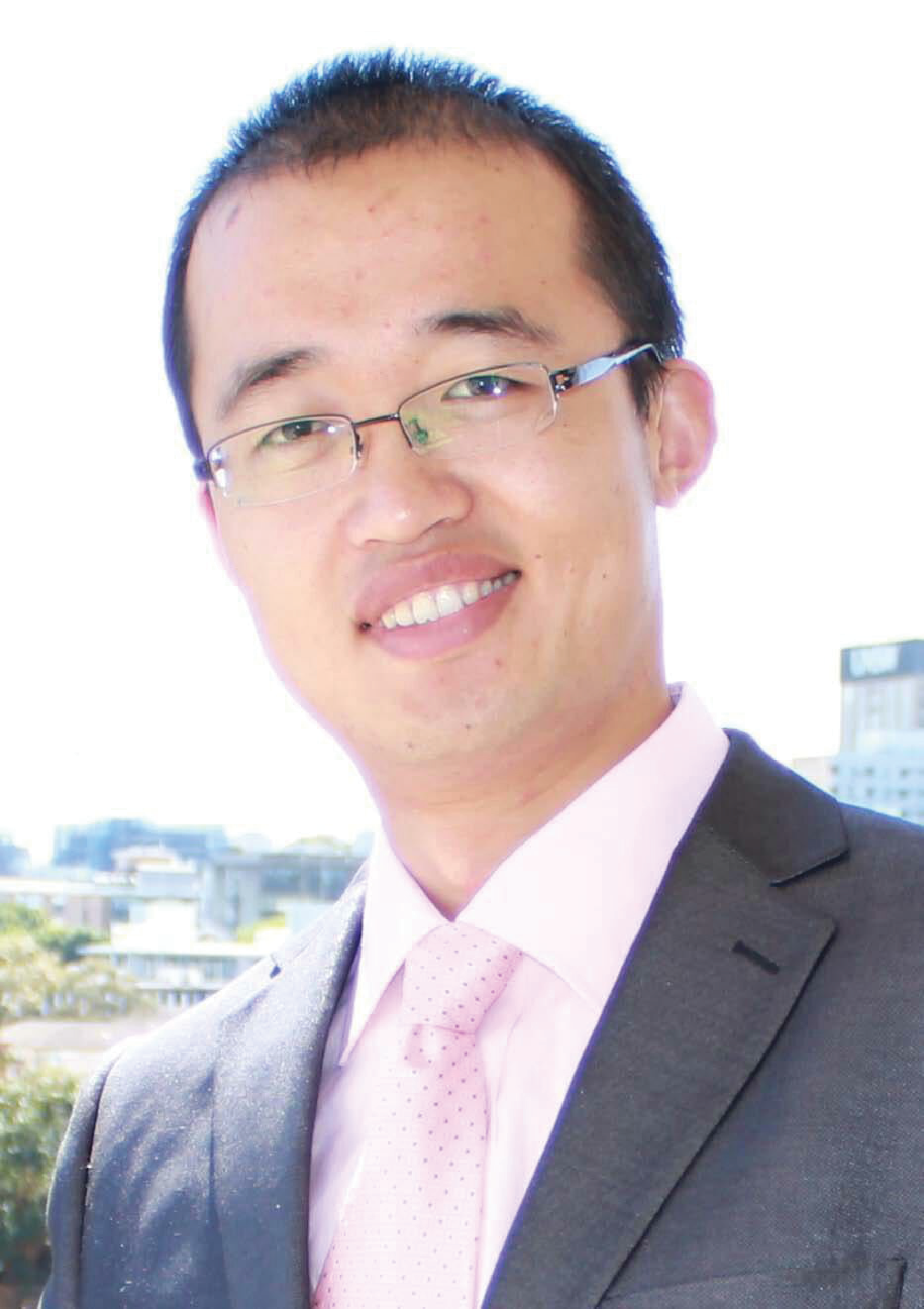}}]{Shihao Yan}
(S'11-M'15) received the Ph.D degree in Electrical Engineering from The University of New South Wales, Sydney, Australia, in 2015. He received the B.S. in Communication Engineering and the M.S. in Communication and Information Systems from Shandong University, Jinan, China, in 2009 and 2012, respectively. From 2015 to 2017, he was a Postdoctoral Research Fellow in the Research School of Engineering, The Australia National University, Canberra, Australia. He is currently a University Research Fellow in the School of Engineering, Macquarie University, Sydney, Australia.
His current research interests are in the areas of wireless communications and statistical signal processing, including physical layer security, covert communications, and location spoofing detection.
\end{IEEEbiography}
\begin{IEEEbiography}[{\includegraphics[width=1in,height=1.25in,clip,keepaspectratio]{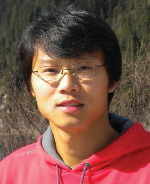}}]{Xiangyun Zhou}
(SM'17) is a Senior Lecturer at the Australian National University (ANU). He received the Ph.D. degree from ANU in 2010. His research interests are in the fields of communication theory and wireless networks. He has been serving as an Editor for various IEEE journals, including IEEE TRANSACTIONS ON WIRELESS COMMUNICATIONS, IEEE WIRELESS COMMUNICATIONS LETTERS and IEEE COMMUNICATIONS LETTERS. He served as a guest editor for IEEE COMMUNICATIONS MAGAZINE's feature topic on wireless physical layer security in 2015. He also served as symposium/track and workshop co-chairs for major IEEE conferences. He was the chair of the ACT Chapter of the IEEE Communications Society and Signal Processing Society from 2013 to 2014. He is a recipient of the Best Paper Award at ICC'11 and IEEE ComSoc Asia-Pacific Outstanding Paper Award in 2016. He was named the Best Young Researcher in the Asia-Pacific Region in 2017 by IEEE ComSoc Asia-Pacific Board.
\end{IEEEbiography}
\begin{IEEEbiography}[{\includegraphics[width=1in,height=1.25in,clip,keepaspectratio]{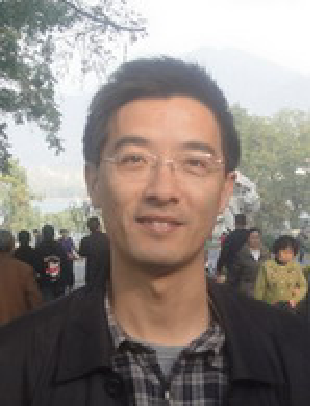}}]{Feng Shu}
was born in 1973. He received the Ph.D., M.S., and B.S. degrees from the Southeast University, Nanjing, in 2002, XiDian University, Xian, China, in 1997, and Fuyang teaching College, Fuyang, China, in 1994, respectively. From Sept. 2009 to Sept. 2010, he is a visiting post-doctor at the University of Texas at Dallas. In October 2005, he joined the School of Electronic and Optical Engineering, Nanjing University of Science and Technology, Nanjing, China, where he is currently a Professor and supervisor of Ph.D and graduate students. He is also with Fujian Agriculture and Forestry University and awarded with Mingjian Scholar Chair Professor in Fujian Province. His research interests include wireless networks, wireless location, and array signal processing. He has published about 200 papers, of which more than 100 are in archival journals including more than 40 papers on IEEE Journals and more than 70 SCI-indexed papers. He holds six Chinese patents.
\end{IEEEbiography}
\begin{IEEEbiography}[{\includegraphics[width=1in,height=1.25in,clip,keepaspectratio]{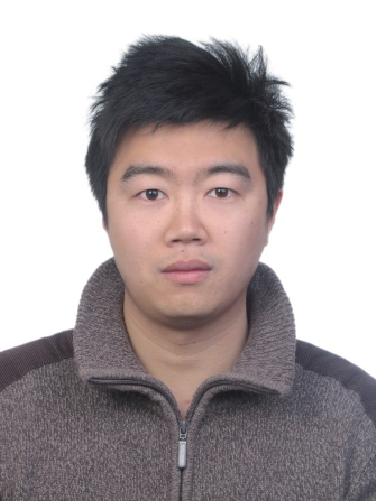}}]{Jun  Li}
(M'09-SM'16) received the Ph.D. degree in electronic engineering from Shanghai Jiao Tong University, Shanghai, China, in 2009. From January 2009 to June 2009, he was with the Department of Research and Innovation, Alcatel Lucent Shanghai Bell, as a Research Scientist. Since 2015, he is with the School of Electronic and Optical Engineering, Nanjing University of Science and Technology, Nanjing, China. His research interests include network information theory, channel coding theory, wireless network coding, and cooperative communications.
\end{IEEEbiography}
\begin{IEEEbiography}[{\includegraphics[width=1in,height=1.25in,clip,keepaspectratio]{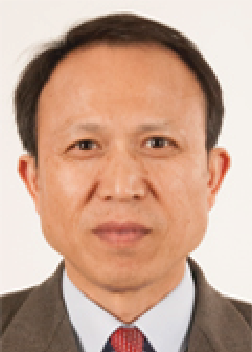}}]{Jiangzhou Wang}
(F'17) is currently Head of the School of Engineering and Digital Arts and a Professor at the University of Kent, United Kingdom. He has authored over 300 papers in international journals and conferences in the areas of wireless mobile communications and three books. He is an IEEE Fellow and IET Fellow. He received the Best Paper Award from IEEE GLOBECOM2012 and was an IEEE Distinguished Lecturer from 2013 to 2014. He is the Technical Program Chair of IEEE ICC2019 in Shanghai. He was the Executive Chair of IEEE ICC2015 in London and the Technical Program Chair of IEEE WCNC2013. He was an Editor for IEEE Transactions on Communications from 1998 to 2013 and was a Guest Editor for IEEE Journal on Selected Areas in Communications, IEEE Communications Magazine, and IEEE Wireless Communications. His research interests include massive MIMO, Cloud RAN, NOMA, D2D, and secure communications.
\end{IEEEbiography}

\end{document}